\documentclass[final, 5p, twocolumn]{elsarticle}

\usepackage[utf8]{inputenc}
\usepackage[T1]{fontenc}
\usepackage[english]{babel}
\usepackage[none]{hyphenat}

\usepackage{amsmath,amssymb,amsfonts}
\usepackage{amstext, mathrsfs, textcomp}
\usepackage{array}

\usepackage{nicefrac}
\usepackage{multirow}

\usepackage{graphicx}
\usepackage[colorlinks=true, citecolor=blue, urlcolor=blue, linkcolor=blue]{hyperref}

\usepackage{xcolor}
\usepackage[export]{adjustbox}

\usepackage{listings}
\definecolor{codegreen}{rgb}{0,0.6,0}
\definecolor{codegray}{rgb}{0.5,0.5,0.5}
\definecolor{codepurple}{rgb}{0.58,0,0.82}
\definecolor{backcolour}{rgb}{0.95,0.95,0.92}
 
\lstdefinestyle{mystyle}{
    backgroundcolor=\color{backcolour},   
    commentstyle=\color{codegreen},
    keywordstyle=\color{blue},
    numberstyle=\tiny\color{codegray},
    stringstyle=\color{codepurple},
    basicstyle=\ttfamily\scriptsize,
    breakatwhitespace=false,         
    breaklines=true,                 
    captionpos=b,                    
    keepspaces=true,                 
    numbers=left,                    
    numbersep=5pt,                  
    showspaces=false,                
    showstringspaces=false,
    showtabs=false,                  
    tabsize=2
}
\DeclareFontFamily{U}{mathb}{\hyphenchar\font45}
\DeclareFontShape{U}{mathb}{m}{n}{<5> <6> <7> <8> <9> <10> gen * mathb
<10.95> mathb10 <12> <14.4> <17.28> <20.74> <24.88> mathb12}{}
\DeclareSymbolFont{mathb}{U}{mathb}{m}{n}
\DeclareMathSymbol{\rcirclearrow}{0}{mathb}{'367}

\usepackage{changes}

\graphicspath{{Figures/}{"Figures/01_GDM/"}{"Figures/02_simulation/"}{"Figures/03_tools/"}{"Figures/04_examples/"}{"Figures/05_technical/"}}

\newcommand{\pygdm}{pyGDM}
\newcommand{\pyGDM}{pyGDM}

\definecolor{pansypurple}{rgb}{0.47, 0.09, 0.29}
\definecolor{lincolngreen}{rgb}{0.11, 0.35, 0.02}
\definecolor{internationalkleinblue}{rgb}{0.0, 0.18, 0.65}

\newcommand{\function}[1]{\textcolor{internationalkleinblue}{\textbf{#1}}}
\newcommand{\object}[1]{\textcolor{lincolngreen}{\textbf{#1}}}

\newcommand{\iu}{\mathrm{i}}
\newcommand{\e}{\mathrm{e}}
\newcommand{\stepsize}{d}
\newcommand{\pdiff}[3][\empty]{\ifx\empty#1
		\frac{\partial\,#2}{\partial #3}
	\else
		\frac{\partial^{#1}\,#2}{\partial #3^{#1}}
	\fi}		

\newcommand*{\functiondescription}[4]{%
  \vspace{0.5\baselineskip}
  
  \noindent\hspace{0.035\linewidth}
  \fbox{
   \begin{minipage}{0.85\linewidth}
    \vspace{0.25\baselineskip}
    \begin{center}
      \ifthenelse{\equal{#1}{f}}
        {\function{#2}\\\small(function)}
        {\object{#2}\\\small(class)}
    \end{center}
    \if\relax\detokenize{#4}\relax
    \else
	\relax
	\vspace{-0.5\baselineskip}\noindent
	\ifthenelse{\equal{#1}{f}}
	    {\textit{arguments:}}
	    {\textit{constructor arguments:}}
	\relax
	\functiondescr{#4}
    \fi
   \end{minipage}
  }
  \vspace{\baselineskip}
  
  \par\noindent\relax
}
\newcommand*{\functiondescr}[1]{%
  \begin{itemize}
   \funcparameter#1\relax
  \end{itemize}
}
\newcommand{\funcparameter}[1]{%
  \ifx\relax#1\empty
  \else
     \vspace{-0.5\baselineskip}
     \item #1
    \relax
    \expandafter\funcparameter
  \fi
}

\newcommand{\TITLE}{\pygdm\ -- A python toolkit for full-field electro-dynamical simulations and evolutionary optimization of nanostructures}

\hypersetup{
    colorlinks,       
    linkcolor=blue,          
    citecolor=blue,         
    filecolor=blue,       
    urlcolor=blue,           
    pdfauthor     = {Wiecha et al.},
    pdftitle      = {\TITLE},
    pdfsubject    = {pyGDM},
    pdfcreator    = {pdflatex},
    pdfkeywords   = {}
    }

\usepackage{etoolbox}
\newtoggle{revtex}
\togglefalse{revtex}

\iftoggle{revtex}
{
	\bibliographystyle{apsrev4-1}}
{
	\bibliographystyle{elsarticle-num}
	\newcommand{\onlinecite}{\cite}
	\newcounter{bla}

	\journal{Computer Physics Communications}
	
}

\begin{document}

\iftoggle{revtex}
{
	\title{\TITLE}
	
	\author{\firstname{Peter R.} \surname{Wiecha}}
	\email[e-mail~: ]{peter.wiecha@cemes.fr}
	\affiliation{CEMES-CNRS, Universit\'e de Toulouse, CNRS, UPS, Toulouse, France}
}
{
	\begin{frontmatter}
		
		
		
		\title{\TITLE}
		
		
		\author[a]{Peter R. Wiecha\corref{author}}
		\cortext[author] {Corresponding author.\\\textit{E-mail address:} peter.wiecha@cemes.fr}
		\address[a]{CEMES-CNRS, Universit\'e de Toulouse, CNRS, UPS, Toulouse, France}
}
\begin{abstract}
%
\pygdm\ is a python toolkit for electro-dynamical simulations in nano-optics based on the Green Dyadic Method (GDM).
In contrast to most other coupled-dipole codes, \pygdm\ uses a generalized propagator, which allows to cost-efficiently solve large monochromatic problems such as polarization-resolved calculations or raster-scan simulations with a focused beam or a quantum-emitter probe.
A further peculiarity of this software is the possibility to very easily solve 3D problems including a dielectric or metallic substrate.
Furthermore, \pygdm\ includes tools to easily derive several physical quantities such as far-field patterns, extinction and scattering cross-section, the electric and magnetic near-field in the vicinity of the structure, the decay rate of quantum emitters and the LDOS or the heat deposited inside a nanoparticle.
Finally, \pygdm\ provides a toolkit for efficient evolutionary optimization of nanoparticle geometries in order to maximize (or minimize) optical properties such as a scattering at selected resonance wavelengths.
\end{abstract}
\iftoggle{revtex}
{\maketitle}
{
	
\end{frontmatter}

\textbf{PROGRAM SUMMARY}

\begin{small}
	\noindent
	\textit{Program Title:} pyGDM                                    \\
	\textit{Licensing provisions:} GPLv3                             \\
	\textit{Programming language:} python, fortran                   \\
	\textit{Nature of problem:}\\
	Full-field electrodynamical simulations of photonic nanostructures. This includes problems like optical scattering, the calculation of the near-field distribution or the interaction of quantum emitters with nanostructures. 
	The program includes a module for automated evolutionary optimization of nanostructure geometries with respect to a specific optical response. \\
	\textit{Solution method:}\\
	The optical response of photonic nanostructures is calculated using field susceptibilities (``Green Dyadic Method'') via a volume discretization. The approach is formally very similar to the coupled dipole approximation.
	\textit{Additional comments including Restrictions and Unusual features (approx. 50-250 words):}\\
	Only 3D nanostructures. The volume discretization is limited to about 10000 meshpoints.
	\\
\end{small}
}




Full-field electro-dynamical simulations are used in nano-optics to predict the optical response of small (often sub-wavelength) particles by solving the Maxwell's equations~\cite{maxwell_dynamical_1865}. 
Examples are either the scattering or the confinement of an external electro-magnetic field by dielectric~\cite{kuznetsov_optically_2016} or metallic~\cite{bharadwaj_optical_2009} nano-structures, the appearance of localized surface plasmons~\cite{maier_plasmonics_2010} or the interaction of nano-structures with quantum emitters placed in their vicinity~\cite{girard_molecular_1995}. 
Nano-optics governs manifold effects and applications.
Examples are phase control or polarization conversion either at the single particle level~\cite{black_optimal_2014, wiecha_polarization_2017} or from metasurfaces~\cite{arbabi_dielectric_2015}, shaping of the directionality of scattering~\cite{curto_unidirectional_2010}, thermoplasmonic heat generation with sub-micrometer heat localization~\cite{baffou_thermo-plasmonics_2013} or nonlinear nano-optics~\cite{kauranen_nonlinear_2012, butet_optical_2015}.
It is of great importance to be able to calculate optical effects occurring in sub-wavelength small structures in order to predict or to interpret experimental findings.

In this paper, we present the python toolkit ``\pygdm'' for full-field electro-dynamical simulations of nano-structures.
Below we list the key-features and aims of \pygdm\ which we will explain in detail in the following.
\begin{itemize}
 \item Easy to use. Easy to install: Fully relying on freely available open-source python libraries (numpy, scipy, matplotlib).
 \item Fast: Performance-critical parts are implemented in fortran and are parallelized with openmp. Efficient and parallelized scipy libraries are used whenever possible. Spectra can be calculated very rapidly via an MPI-parallelized routine.
 \item Electro-dynamical simulations including a substrate.
 \item Different illumination sources such as plane wave, focused beam or dipolar emitter.
 \item Efficient calculation of large problems such as raster-scan simulations.
 \item Provide tools to rapidly post-process the simulations and derive physical quantities such as
  \begin{itemize}
   \item optical near-field inside and around nanostructures.
   \item extinction, absorption and scattering cross-sections.
   \item polarization- and spatially resolved far-field scattering.
   \item heat generation, temperature distribution around nano-objects.
   \item photonic local density of states (LDOS).
   \item modification of the decay-rate of dipolar emitters in the presence of a nanostructure.
  \end{itemize}
 \item Evolutionary optimization of the nano-particle geometry with regards to specific optical properties.
 \item Easy to use visualization tools including animations of the electro-magnetic fields.
\end{itemize}

We will start with a brief introduction to the Green Dyadic Method (GDM), the numerical discretization scheme and the renormalization of the Green's dyad. 
We will also compare the GDM to other frequently used numerical techniques.
In the second part, we will explain in more detail the main features and tools provided by \pygdm.
We start by explaining the general structure of \pyGDM\ and the ingredients to setup a simulation. 
Then we describe how the main simulation routines work. 
This part is followed by descriptions of the \pyGDM-tools for simulating different optical effects, post-processing, data-analysis and visualization.
Subsequently we will illustrate the capabilities of \pygdm\ by some example simulations and benchmarks. In particular, we will compare \pygdm-simulations to Mie theory.
Finally, we will give an overview on the evolutionary optimization submodule of \pygdm, accompanied by several examples.
In the appendix we provide details concerning more technical tools and aspects of \pygdm\ as well as instructions for the compilation, installation and use of \pygdm.

\section{The Green dyadic method}\label{sec:green_method}

In the following we will give a brief introduction to the basic concepts of the Green dyadic method, implemented in \pygdm. 
Before we begin with this short overview, we want to note that the GDM is a frequency domain technique, solving Maxwell's equations for mono\-chroma\-tic fields (oscillating at fixed frequency~\(\omega\)).
\paragraph*{Note:} We use cgs (centimeter, gram, second) units in \pygdm\ which results in simpler terms for most of the equations. 
This is first of all helpful for the derivation of the main equations and has no impact on the simulation results. 
The post-processing routines return values conform with SI units such as cross-sections (units of nm\(^2\)), powers in Watt or unit-less values (e.g. relative field intensities such as \(|\mathbf{E}|^2 / |\mathbf{E}_0|^2\)).

\subsection{From Maxwell's equations to Lippmann-Schwinger equation}

\begin{figure}[t]
  \centering
  \includegraphics[width=.65\columnwidth]{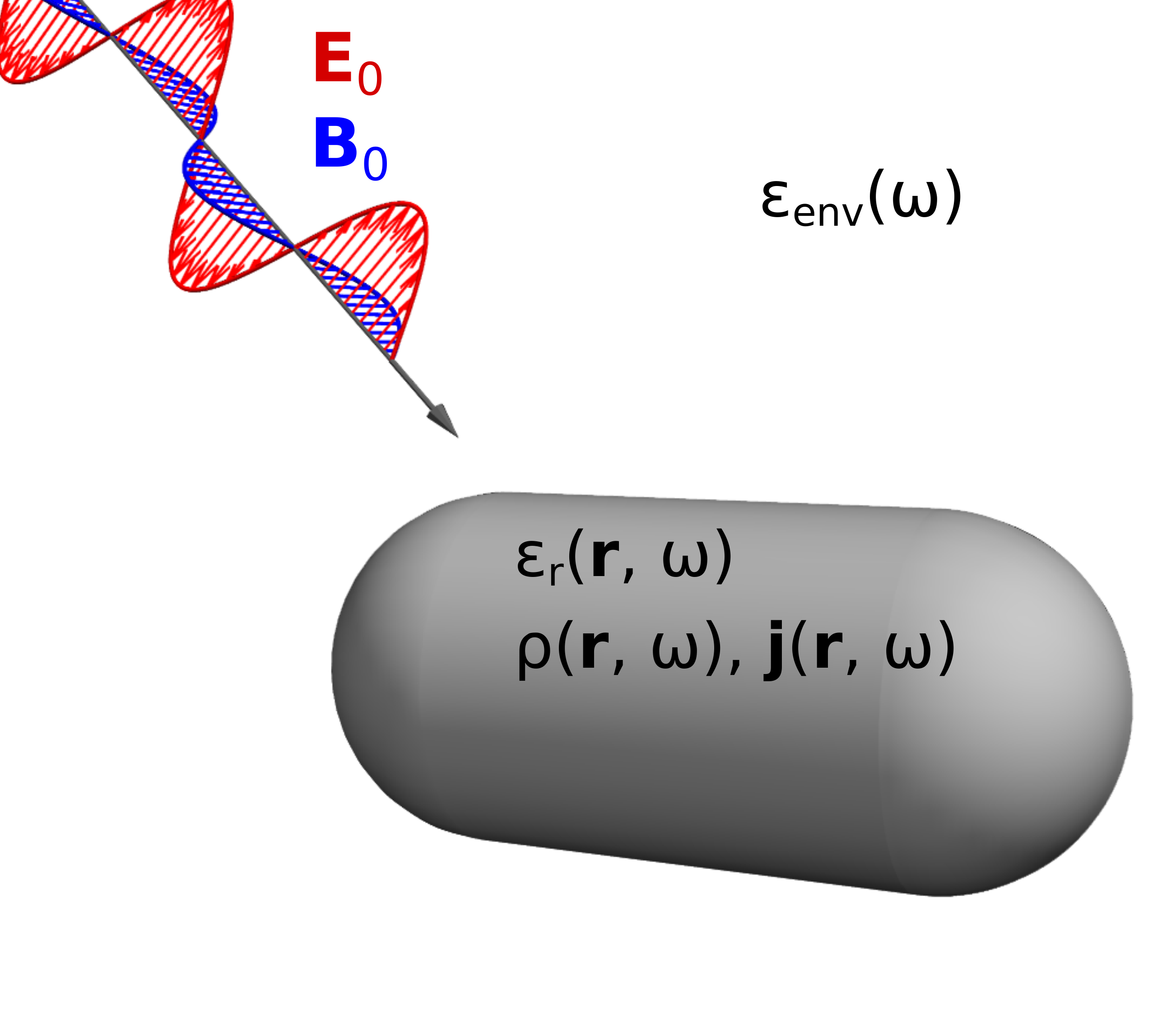}
  \caption{
  Electromagnetic wave impinging on a nano\-structure of arbitrary shape, placed in a homogeneous environment.}\label{fig:Theory_sketch_object}
\end{figure}

All electromagnetic phenomena can be entirely described by the four Maxwell's equations, which in the frequency domain write as follows (cgs units):
%
\begin{subequations}\label{eq:Maxwell}
  \begin{align}
  \nabla\cdot \mathbf{E}(\mathbf{r}, \omega)          &= \frac{4\pi}{\epsilon_{\text{env}}} \rho(\mathbf{r}, \omega)         \label{eq:MaxwellFourierdivD}\\
  \nabla \times \mathbf{E}(\mathbf{r}, \omega)          &= \iu k_0 \mathbf{B}(\mathbf{r}, \omega)          \label{eq:MaxwellFourierrotE}\\
  \nabla\cdot \mathbf{B}(\mathbf{r}, \omega)          &= 0            \label{eq:MaxwellFourierdivB}\\
  \nabla\times \mathbf{B}(\mathbf{r}, \omega)          &= -\iu k_0 \epsilon_{\text{env}} \mathbf{E}(\mathbf{r}, \omega) + \frac{4\pi}{c} \mathbf{j}(\mathbf{r}, \omega)    \label{eq:MaxwellFourierrotH}
  \end{align}
\end{subequations}
where the charge density \(\rho\) and the current density \(\mathbf{j}\) are associated with an arbitrary nanostructure, placed in an environment of permittivity \(\epsilon_{\text{env}}\) (\textit{c.f.} Fig.~\ref{fig:Theory_sketch_object}).
\(k_0 = \omega / c\) is the wavenumber of light in vacuum, \(c\) the speed of light and the symbol \(\times\) is the rotational.
%
%
%
\(\epsilon_r\) and \(\mu_r\) are the relative dielectric permittivity and magnetic permeability of the nanostructure, respectively. 
For dispersive media, \(\epsilon_r\) and \(\mu_r\) are functions of the frequency \(\omega\).
They are defined as the ratios of the material's permittivity and permeability relative to the vacuum values \(\epsilon_0\) and \(\mu_0\). 
They can be related to the electric and magnetic susceptibilities as \(\chi_{\text{e}} = (\epsilon_r - \epsilon_{\text{env}})/4\pi\) and \(\chi_{\text{m}} = (\mu_r - \mu_{\text{env}})/4\pi\), respectively.
In general, \(\chi_{\text{e}} (\mathbf{r}, \omega)\) and  \(\chi_{\text{m}} (\mathbf{r}, \omega)\) are functions of frequency and space.
In \pygdm\ we assume non-magnetic media, hence \(\mu_r = \mu_{\text{env}} = 1\).

%
%

It is possible to derive a wave-equation for the electric field from Maxwell's equations (see e.g. Ref.~\onlinecite{griffiths_introduction_1989}, chapter~9 or Ref.~\onlinecite{girard_near_2005}):
\begin{equation}
  (\Delta + k^2) \mathbf{E}(\mathbf{r}, \omega) = 
      - \frac{4\pi}{\epsilon_{\text{env}}}
        \left( k^2  + \nabla \nabla \right) 
        \mathbf{P}(\mathbf{r}, \omega).
\label{eq:waveequationEfieldSI}
\end{equation}
Where \(\nabla\) and \(\Delta\) are the nabla- and Laplace operator, respectively, \(\mathbf{P} = \boldsymbol\chi_{\text{e}} \cdot \mathbf{E}\) is the electric polarization and \(k\) the wavenumber in the environment medium with \(k = \sqrt{\epsilon_{\text{env}}}\,k_0\).

\paragraph*{Note:}
The dielectric function is in general a tensor of rank 2. 
In \pygdm, an isotropic susceptibility \(\chi_{\text{e,iso}}\) is assumed, hence the susceptibility tensor \(\boldsymbol\chi_{\text{e}}\) is defined as
\begin{equation}
\boldsymbol\chi_{\text{e}}(\mathbf{r}, \omega) = 
\left[
\begin{matrix}  
\chi_{\text{e,iso}}(\mathbf{r}, \omega) & 0 & 0 \\ 
0 & \chi_{\text{e,iso}}(\mathbf{r}, \omega) & 0 \\ 
0 & 0 & \chi_{\text{e,iso}}(\mathbf{r}, \omega)
\end{matrix}
\right]\, .
\end{equation}
In future versions of \pygdm\ anisotropic polarizabilities might be supported.

From the wave-equation Eq.~\eqref{eq:waveequationEfieldSI} one can derive a vectorial Lippmann-Schwinger equation for the electric field (see e.g. Ref.~\onlinecite{girard_near_2005}):
\begin{equation}
 \mathbf{E}(\mathbf{r}, \omega)  = 
     \mathbf{E}_0(\mathbf{r}, \omega) + 
         \int \mathbf{G}_{\text{tot}}^{\text{EE}}(\mathbf{r}, \mathbf{r'}, \omega) \cdot 
              \boldsymbol\chi_{\text{e}} \cdot \mathbf{E}(\mathbf{r'}, \omega) \text{d} \mathbf{r'} 
 \label{eq:LippmannSchwingerG0}
\end{equation}
which relates in a self-consistent manner the incident (or ``zero order'', ``fundamental'') electric field \(\mathbf{E}_0\) with the total field \(\mathbf{E}\) inside the structure of susceptibility \(\boldsymbol\chi_{\text{e}}\). 
The integral in Eq.~\eqref{eq:LippmannSchwingerG0} runs over the volume of the structure.
\(\mathbf{G}_{\text{tot}}^{\text{EE}}\) is the Green's dyad, describing the environment in which the structure is placed (see also section~\ref{sec:technical_detail}).
The Green's dyadic tensors \(\mathbf{G}\) are also called field susceptibilities and were originally introduced by G. S. Agarwal. \cite{agarwal_quantum_1975}. For an object in vacuum \(\mathbf{G}_{\text{tot}}^{\text{EE}}=\mathbf{G}_0^{\text{EE}}\), which writes~\cite{girard_near_2005, girard_shaping_2008}
\begin{multline}\label{eq:vacuumGreenDyadicFunction}
 \mathbf{G}_0^{\text{EE}}(\mathbf{r}, \mathbf{r'}, \omega) = 
    \frac{1}{\epsilon_{\text{env}}} 
    \Big( k^2 \, \mathbf{I} + \nabla\nabla \Big) 
    G_0(\mathbf{r}, \mathbf{r'}, \omega) \\
   = \frac{ \e^{\iu k R} }{ \epsilon_{\text{env}} } \,
   \Big( - k^2 \mathbf{T}_1(\mathbf{R})
         - ik \mathbf{T}_2(\mathbf{R}) 
         + \mathbf{T}_3(\mathbf{R})     \Big).
\end{multline}
\(\mathbf{I}\) is the Cartesian unitary tensor, \(\nabla\) the nabla operator acting along \(\mathbf{r}\) and \(G_0\) the scalar Green's function (see equation~\eqref{eq:scalarGreenFunctionVac}). The superscript ``\(^{\text{EE}}\)'' indicates that the Green's function accounts for an electric-electric interaction.
Furthermore we used the abbreviations \(\mathbf{R} = \mathbf{r} - \mathbf{r'}\) and
\begin{align}
 \mathbf{T}_1(\mathbf{R}) & = \frac{\mathbf{R}\mathbf{R} - \mathbf{I}R^2}{R^3} \label{eq:vacuumGreenDyadicFunctionT1}\\
 \mathbf{T}_2(\mathbf{R}) & = \frac{3\mathbf{R}\mathbf{R} - \mathbf{I}R^2}{R^4} \label{eq:vacuumGreenDyadicFunctionT2} \\
 \mathbf{T}_3(\mathbf{R}) & = \frac{3\mathbf{R}\mathbf{R} - \mathbf{I}R^2}{R^5}. \label{eq:vacuumGreenDyadicFunctionT3}
\end{align}
\(\mathbf{R}\mathbf{R}\) is the tensorial product of \(\mathbf{R}\) with itself and \(R\) represents its modulus.
\(\mathbf{T}_1\) describes far-field effects while \(\mathbf{T}_2\) and \(\mathbf{T}_3\) account for the near-field.

\begin{figure}[tp]
  	\centering
  	\includegraphics[width=\linewidth]{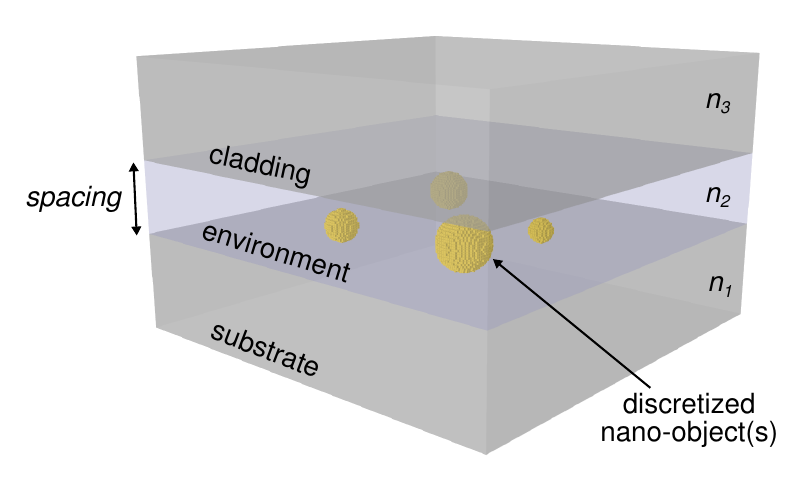}
  	\caption{
  		Geometry of the reference system described by the Green's dyad used in \pygdm:
  		The discretized nano-structure is placed in the environment layer with (complex) refractive index \(n_2\) and of thickness \textit{spacing}. 
  		It is sandwiched between a substrate (\(n_1\)) and a cladding layer (\(n_3\)).
  	}\label{fig:reference_system}
  \end{figure}

In \pygdm\ an additional non-retarded Green's dyad is used which allows to include a substrate and a cladding layer (see Fig.~\ref{fig:reference_system}):
\begin{equation}
 \mathbf{G}_{\text{tot}}^{\text{EE}} =  \mathbf{G}_0^{\text{EE}} + \mathbf{G}_{\text{3-layer}}\, .
\end{equation}
Such dyadic function \(\mathbf{G}_{\text{3-layer}}\) for a layered reference system can be derived in an asymptotic form using the image charges method (see also section~\ref{sec:technical_detail}).
The derivation of a retarded Green's dyad for multi-layered systems is explained in detail e.g. in Refs.~\onlinecite{colas_des_francs_enhanced_2005, paulus_accurate_2000}.
For a derivation of the Lippmann-Schwinger equation in SI units, see e.g. Ref.~\onlinecite{wiecha_linear_2016}.

  %


\subsection{Volume discretization}

\begin{figure*}[t]
  \centering
  \includegraphics[width=.4\textwidth]{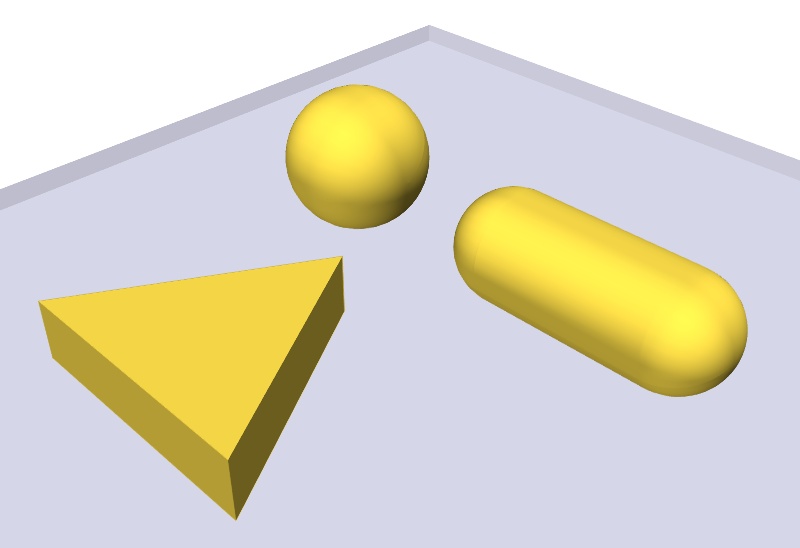}
 \hspace{1cm}
  \includegraphics[width=.4\textwidth]{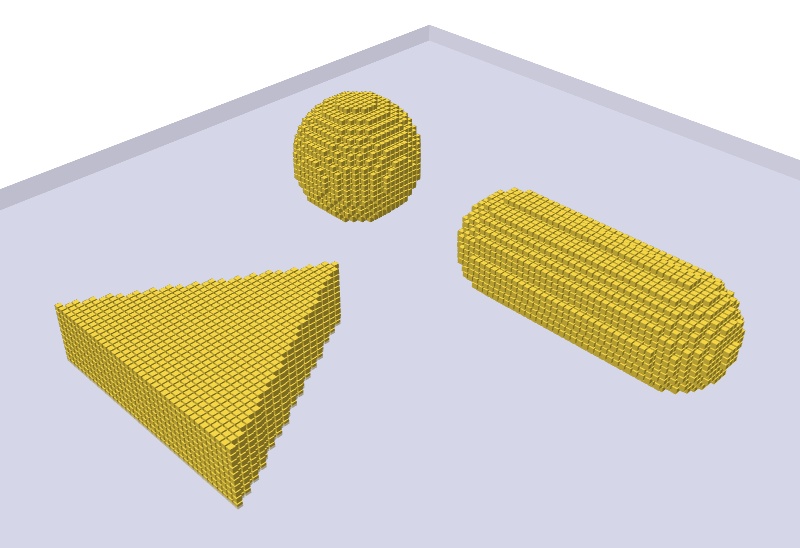}
  \caption{
  Arbitrary nanostructure composed of multiple elements lying on a substrate (left) and its volume discretization on a cubic lattice (right).}\label{fig:Theory_Volume_Discretization}
\end{figure*}

For arbitrarily shaped objects, the integral in the Lippmann-Schwinger equation~\eqref{eq:LippmannSchwingerG0} can generally not be solved analytically.
In the following we describe a numerical approach which requires the discretization of the integral into a sum over finite size volume elements (see also Ref.~\onlinecite{girard_near_2005}).
For reasons of clarity the dependency on the frequency \(\omega\) will be omitted in the following.
We discretize the nano-object using \(N\) cubic volume elements centered at positions \(\mathbf{r}_i\), as illustrated in figure~\ref{fig:Theory_Volume_Discretization}. 
The cube side lengths \(\stepsize\) and thus \(V_{\text{cell}} = \stepsize^3\) are constant on the mesh.
\begin{multline}
 \mathbf{E}(\mathbf{r}_i, \omega) = 
    \mathbf{E}_0(\mathbf{r}_i, \omega) + \\
    \sum\limits_{j=1}^{N} 
      \mathbf{G}_{\text{tot}}^{\text{EE}}(\mathbf{r}_i, \mathbf{r}_j, \omega) \cdot 
      \boldsymbol\chi_{\text{e}}(\mathbf{r}_j,\omega)\cdot \mathbf{E}(\mathbf{r}_j, \omega) V_{\text{cell}}.
  \label{eq:LippmannSchwingerVolumeDiscretization}
\end{multline}
%
%
We can rewrite eq.~\eqref{eq:LippmannSchwingerVolumeDiscretization} as follows
\begin{multline}
    \mathbf{E}_0(\mathbf{r}_i) 
       = \mathbf{E}(\mathbf{r}_i) - 
      \sum\limits_{j=1}^{N} 
	\mathbf{G}_{\text{tot}}^{\text{EE}}(\mathbf{r}_i, \mathbf{r}_j) \cdot 
	\boldsymbol\chi_{\text{e}}(\mathbf{r}_j)\cdot \mathbf{E}(\mathbf{r}_j) V_{\text{cell}} \\
       = \sum\limits_{j=1}^{N} 
	\Big( \delta_{ij} \mathbf{I} -
	    \boldsymbol\chi_{\text{e}}(\mathbf{r}_j) \cdot V_{\text{cell}} \, \mathbf{G}_{\text{tot}}^{\text{EE}}(\mathbf{r}_i, \mathbf{r}_j)
	\Big)
	 \cdot \mathbf{E}(\mathbf{r}_j)
\end{multline}
where \(\delta_{ij}\) is the Kronecker symbol.

Let us now define two \(3N\)-dimensional vectors containing the ensemble of all electric field vectors in the discretized nano-object
\begin{multline*}
  \mathbf{E}_{0, \text{obj.}} = 
   \Big( 
     E_{0,x}(\mathbf{r}_1),E_{0,y}(\mathbf{r}_1),E_{0,z}(\mathbf{r}_1), \\
     E_{0,x}(\mathbf{r}_2),\;  \ldots,\quad 
     \ldots,\; E_{0,z}(\mathbf{r}_N)
   \Big) 
\end{multline*}
\begin{multline*}
  \mathbf{E}_{\text{obj.}} = 
   \Big( 
     E_{x}(\mathbf{r}_1),E_{y}(\mathbf{r}_1),E_{z}(\mathbf{r}_1), \\
     E_{x}(\mathbf{r}_2),\;  \ldots,\quad 
     \ldots,\; E_{z}(\mathbf{r}_N)
   \Big). 
\end{multline*}
Together with the \(3N \times 3N\) matrix \(\mathbf{M}\) composed of \(3 \times 3\) sub-matrices
\begin{equation}\label{eq:definitionMforInversion}
 \mathbf{M}_{ij} = \delta_{ij} \mathbf{I} - \boldsymbol\chi_{\text{e}} (\mathbf{r}_j) \cdot V_{\text{cell}} \mathbf{G}_{\text{tot}}^{\text{EE}}(\mathbf{r}_i, \mathbf{r}_j)
\end{equation}
we obtain a coupled system of \(3N\) linear equations
\begin{equation}\label{eq:definitionMainEquationGDM}
 \mathbf{E}_{0, \text{obj.}} = \mathbf{M} \cdot \mathbf{E}_{\text{obj.}}\,.
\end{equation}
If we inverse the matrix~\(\mathbf{M}\) defined by eq.~\eqref{eq:definitionMforInversion}, we can calculate the field \(\mathbf{E}_{\text{obj.}}\) inside the structure for all possible incident fields \(\mathbf{E}_{0, \text{obj.}}\) (at frequency \(\omega\)) by means of a simple matrix-vector multiplication:
\begin{equation}\label{eq:definitionGeneralizedPropoagator}
 \mathbf{E}_{\text{obj.}} = {\boldsymbol{\cal K}} \cdot \mathbf{E}_{0, \text{obj.}}\,,
\end{equation}
where we used the symbol \({\boldsymbol{\cal K}}\) for the inverse matrix
\begin{equation}
 {\boldsymbol{\cal K}}(\omega) = \mathbf{M}^{-1}(\omega)\, .
\end{equation}
\({\boldsymbol{\cal K}}\) is called the \emph{generalized field propagator}, as introduced by Martin \textit{et al.} \cite{martin_generalized_1995}.

\paragraph*{Note:} In our notation, \({\boldsymbol{\cal K}}\) represents the full \(3N \times 3N\) matrix, describing the response of the entire nanostructure. 
This matrix is composed of \(3\times 3\) sub-tensors \(\mathbf{K}(\mathbf{r}_i,\mathbf{r}_j)\) for the couples of \(i\)th and \(j\)th meshpoint.

\paragraph*{Note:} After equation~\eqref{eq:LippmannSchwingerVolumeDiscretization}, we can use the Green's dyad of the reference system with the field inside the particle in order to calculate the total electric field at any point \(\mathbf{r}_i\) outside the nanostructure.

\subsection{Renormalization of the Green's dyad}\label{sec:GDMRenormalizationScheme}

When integrating the polarization distribution in equation~\eqref{eq:LippmannSchwingerG0} over the volume of the nanostructure, we integrate scalar Green's functions of the form
\begin{equation}\label{eq:scalarGreenFunctionVac}
   G_0(\mathbf{r}, \mathbf{r'})
   = \frac{\e^{\iu k\, | \mathbf{r} - \mathbf{r'} |}}{ | \mathbf{r} - \mathbf{r'} | }.
\end{equation}
Obviously, \(G_0\) diverges if \(\mathbf{r} = \mathbf{r'}\), which occurs when the field of a point dipole \(\mathbf{p}\delta(\mathbf{r} - \mathbf{r'})\) is being evaluated at the dipole's position \(\mathbf{r'}\) itself. 
As a consequence, in order to remove this singularity, we need to apply a regularization scheme~\cite{yaghjian_electric_1980}.
For a three dimensional cubic mesh, a simple renormalization rule for the free-space Green's dyad has been proposed (see Ref.~\onlinecite{girard_near-field_1996}, section~4.3):
\begin{equation}\label{eq:renormalization_cube}
 \mathbf{G}_{0,\text{cube}}^{\text{EE}}(\mathbf{r}_i, \mathbf{r}_i) = 
 - \frac{4\pi}{3 \epsilon_{\text{env}} \stepsize^3} \, \mathbf{I}
\end{equation}
with \(\stepsize\) the stepsize of the volume discretization.

The choice of an appropriate mesh can be crucial for the convergence of the method. 
While structures with flat surfaces and right angles (e.g. cuboids) can be accurately discretized using a cubic mesh, particles with 3-fold symmetry (e.g. prisms) or curved structures like wires of circular section or spherical particles are better described using a hexagonal mesh. 
A \(3\)D hexagonal compact mesh can be regularized with (see Ref.~\onlinecite{girard_shaping_2008}, section~3.1)
\begin{equation}\label{eq:renormalization_hex}
 \mathbf{G}_{0,\text{hex}}^{\text{EE}}(\mathbf{r}_i, \mathbf{r}_i) = 
 -\frac{4\pi \sqrt{2}}{3 \epsilon_{\text{env}} \stepsize^3}  \, \mathbf{I}
 \, .
\end{equation}
While a cubic mesh cell has a volume of \(V_{\text{cell}} = \stepsize^3\), in the hexagonal compact case, the volume of a cell equals \(V_{\text{cell}} = \stepsize^3 / \sqrt{2}\) and also must be accordingly adapted in Eq.~\eqref{eq:definitionMforInversion}.

Other geometries like cuboids~\cite{ould_agha_near-field_2014} or tetrahedrons~\cite{kottmann_accurate_2000} can be used for the mesh as well, but are not implemented in \pygdm\ so far.
Because it accounts for the field of a point dipole at the location of the dipole itself, the sub-matrix \(\mathbf{M}_{ii}\) is also called ``self-term''.


\subsection{Multiple monochromatic simulations on the same nanostructure}\label{sec:GDMRasterScanSimulations}
Once the generalized propagator \(\mathbf{K}\) is known, we can calculate the response of the system to arbitrary monochromatic incident fields (e.g. plane waves, focused beams or even fast electrons) by means of a simple matrix-vector multiplication. 
This can be used for instance to do raster-scan simulations at low numerical cost, by raster-scanning a light source such as a focused incident beam or a dipolar emitter step-by-step over the nano-object, while calculating and eventually post-processing the field at each position~\cite{teulle_scanning_2012}.

\section{Comparison to other electro-dynamical simulation techniques}

Before proceeding with a detailed introduction to the \pygdm\ toolkit, we want to give a non-exhaustive overview of other methods commonly used for solving electro-dynamical problems in nano-optics.

A widely used frequency domain solver is the open source software DDSCAT \cite{draine_discrete-dipole_1994}, which implements a frequency domain technique analog to the GDM.
It is usually called the ``Coupled'' or ``Discrete Dipole Approximation'' (CDA or DDA, respectively). 
However, there exist two main differences to GDM as used in this work. 
First, the renormalization problem is circumvented by setting the self-terms to zero and including the corresponding contributions using a physical polarizability for each dipole. Using such physical polarizabilities (usually of spherical entities) for each mesh-cell however leads generally to a worse convergence for larger step-sizes.
The second difference is more technical. In the DDSCAT implementation of DDA, the matrix \(\mathbf{M}_{\text{DDSCAT}}\) is not stored in memory (c.f. Eq.~\eqref{eq:definitionMforInversion}).
The resolution of the inverse problem is done by the conjugate gradients method, where the elements \(M_{\text{DDSCAT}, ij}\) are computed \textit{on-demand} during the calculation of the vector-matrix products \(\mathbf{M}_{\text{DDSCAT}} \cdot \mathbf{x} = \mathbf{E}\).
To speed up these matrix-vector multiplications, a scheme involving fast Fourier transformations (FFT) is used \cite{goodman_application_1991}.
A drawback is that without storing \(\mathbf{M}\), efficient preconditioning is very difficult (see also appendix~\ref{sec:conjugate_gradients}).
Convergence of the DDSCAT conjugate gradient iterative scheme is therefore relatively slow and only obtained for very fine discretization meshes, further slowing down the computation due to the large size of the coupled dipole matrix \(\mathbf{M}_{\text{DDSCAT}}\).
An obvious advantage of DDSCAT is, that large problems with huge numbers of mesh points can be treated, since the matrix coupling all dipoles is not stored in memory. 
However, the advantage of the generalized propagator is lost.
The calculation of different incident fields at a fixed wavelength (such as raster-scan simulations) requires to re-run the time-consuming conjugate gradients solver for each configuration.
Another free implementation of the DDA with particular focus on electron energy loss spectroscopy (EELS) simulations is the DDEELS package~\cite{geuquet_eels_2010}.

Maxwell's equations can be reformulated as a set of surface-integral equations. 
It is therefore possible to develop a similar formalism as the above explained volume integral method in which only the surfaces of a nanostructure are discretized instead of the volume \cite{garcia_de_abajo_retarded_2002}.
A great advantage of this so-called Boundary Element Method (BEM) is the smaller amount of discretization cells, which however comes at the cost of a more complex mathematical framework and numerical implementation.
With MNPBEM an open-source BEM-implementation for MATLAB exists which allows also the consideration of layered environments~\cite{hohenester_mnpbem_2012, waxenegger_plasmonics_2015}.

\begin{figure*}[t]
  \centering
  \includegraphics{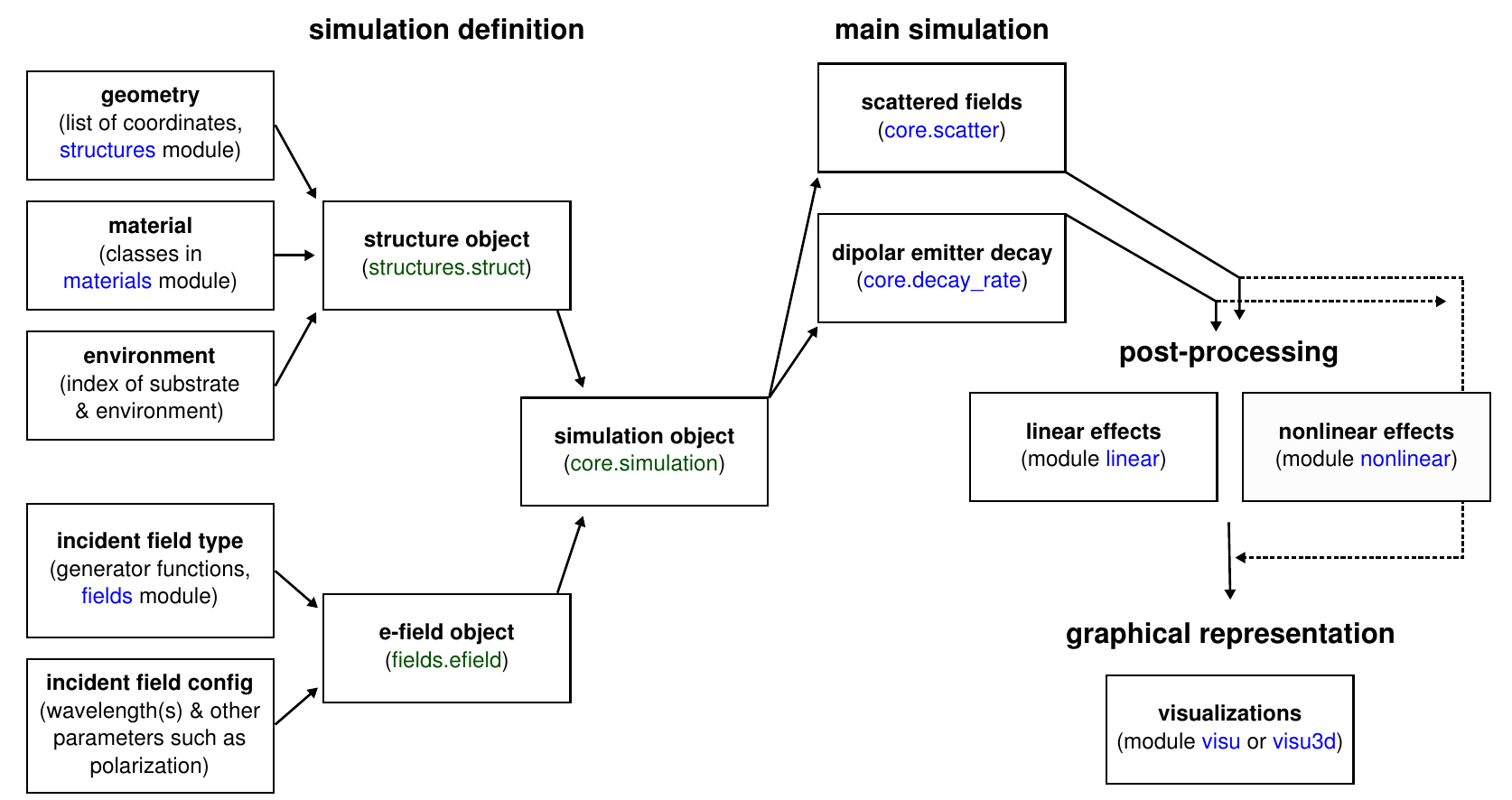}
  \caption{
  Structure of the \pygdm\ package and workflow of a typical simulation:
  (1) Setup of the geometry, environment and incident electric field. This is bundled in an instance of the \object{simulation} object.
  (2) Main GDM simulation.
  (3) Possible post-processing (e.g. calculation of extinction cross-sections).
  (4) Visualization of the results.
  }\label{fig:structure_pygdm}
\end{figure*}

Another very popular and flexible technique for electrodynamical simulations is the Finite-Difference Time-Domain (FDTD) method \cite{inan_numerical_2011, baida_finite_2013, cao_electron_2015}.
As the name suggests, the calculation is performed in the time domain, which means that Maxwell's equations are iteratively evolved by small time increments.
The problem is discretized in both, space and time.
An incoming wave travels time-step by time-step across the region of interest and when the wave-packet has passed or turn-on effects have fully decayed (e.g. for plane wave illumination), the actual numerical measurement is performed.
With respect to computational time, a disadvantage is the additional dimension (time) that needs to be discretized. 
Furthermore, a fraction of the environment around the object of interest has to be included in the discretization space, which is why FDTD is called a ``domain discretization technique''.
Particularly in \(3\)D problems, this can lead to very high computational costs.
Another drawback of FDTD can be the low accuracy for near-field intensities if very strong field enhancements occur (\textit{e.g.} in plasmonics) \cite{hoffmann_comparison_2009}.
However, the simplicity and the robustness of the method are great advantages of FDTD.
Furthermore, using temporally short and therefore spectrally broad illumination pulses, a large frequency spectrum can be obtained in a single simulation run. 
Frequency domain techniques on the other hand require each wavelength to be calculated separately.
Provided an accurate analytical model for the material dispersion exists, this advantage can compensate the larger discretization domain in spectral simulations, compared to frequency domain methods like the GDM.
A powerful open source implementation that comes with a rich toolbox is the software ``MEEP'' \cite{oskooi_meep_2010}. 
For a general introduction on finite difference methods, see for example Ref.~\onlinecite{press_numerical_2007}, chapter~17.

Finally, a very popular domain discretization technique in the frequency domain is the Finite Element Method (FEM, e.g. implemented in the commercial software ``COMSOL Multiphysics'').
Due to its adjustable mesh-size it is particularly apt for plasmonic problems, where extremely localized fields can occur at sharp extremities. 
However, it suffers from the same drawback as FDTD since a certain volume around the nano-object needs to be discretized and included in the calculation, often leading to high memory and CPU-time requirements.

A review including benchmarks for different numerical techniques in nano-optics can be found in Ref.~\onlinecite{smajic_comparison_2009}.
An extensive discussion of different DDA variants including a detailed review on their accuracies is given in Ref.~\onlinecite{yurkin_discrete_2007}.

\section{Setting up a \pygdm\ simulation}\label{sec:setup_simulation}

The structure of the \pygdm\ package and the main steps to setup and run a simulation are schematically depicted in figure~\ref{fig:structure_pygdm}.
The heart of \pygdm\ is the \object{simulation} object which contains the information about the structure, its environment and the incident electro-magnetic field(s) used in the simulation:
\functiondescription{o}{core.simulation}
  { {\textit{struct:} instance of \object{structures.struct}} 
    {\textit{efield:} instance of \object{fields.efield}} } 
  {}
A minimal example of a \pygdm\ python script is provided in section~\ref{sec:technical_detail}.

\subsection{Geometry and material dispersion}

The geometry of the nanostructure and the dielectric constant of both its constituent material and the environment are stored in an instance of
\functiondescription{o}{structures.struct}
  { {\textit{step:} the stepsize (nm)} 
    {\textit{geometry:} the particle geometry as \textbf{list} of meshpoints (\(x,y,z\))} 
    {\textit{material:} an instance of a material object, having a routine \function{epsilon(wavelength)} which returns the permittivity at \textit{wavelength} (\textit{wavelength} in nm).}
    {\textit{n1,n2,n3:} the (complex) refractive indices of (1) the substrate, (2) the particle environment and (3) a top layer} 
    {\textit{normalization:} the normalization factor for the Green's dyad (use \function{structures.\allowbreak get\_\allowbreak normalization})} }
  {}
which contains the geometry as a list of mesh-point coordinates and the material dispersion via an instance of some \object{materials.dispersion\_\allowbreak class}.

\subsection{Excitation fields}

The second key-ingredient of a \pygdm-simulation is the incident (illuminating) electro-magnetic field.

The fields in the GDM are time-harmonic, oscillating at frequency \(\omega\).
We describe these fields using the phasor description with complex amplitudes in which we include the phase information:
\begin{equation}
 \mathbf{\tilde{E}}(\mathbf{r}, \omega, t) = 
 \mathbf{\hat{E}}(\mathbf{r}, \omega)\, \e^{- \iu \omega t}\,\e^{\iu \varphi} = 
 \mathbf{E}(\mathbf{r}, \omega)\, \e^{-\iu  \omega t}.
\label{eq:harmonicComplexWave}
\end{equation}
\(\mathbf{\tilde{E}}\) is the electric field including the time-dependence. 
We assume time-harmonicity, thus the time-dependence is expressed by the term \(\e^{-\iu \omega t}\). 
\(\mathbf{\hat{E}}\) is the real valued amplitude, \(\mathbf{E}\) the complex amplitude (the ``phasor'') which includes the phase-factor \(\e^{\iu \varphi}\) in its imaginary part.

\begin{figure*}[t]
  \centering
  \includegraphics[width=.195\textwidth]{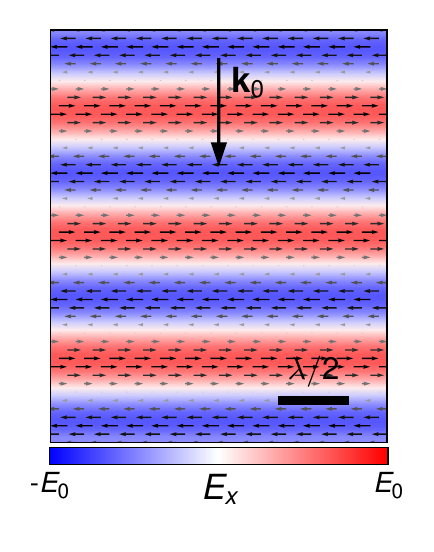}
  \includegraphics[width=.195\textwidth]{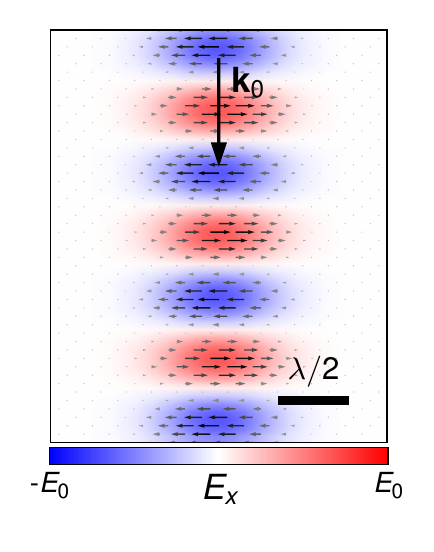}
  \includegraphics[width=.195\textwidth]{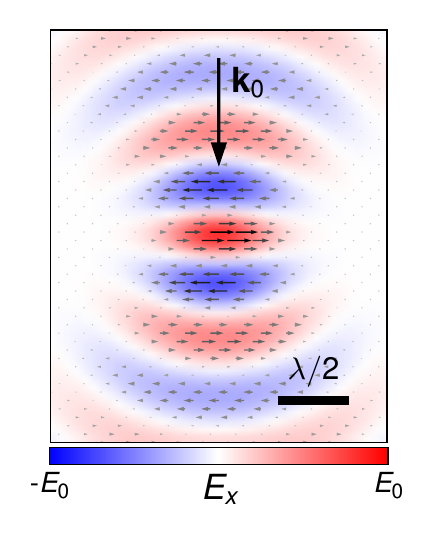}
  \includegraphics[width=.195\textwidth]{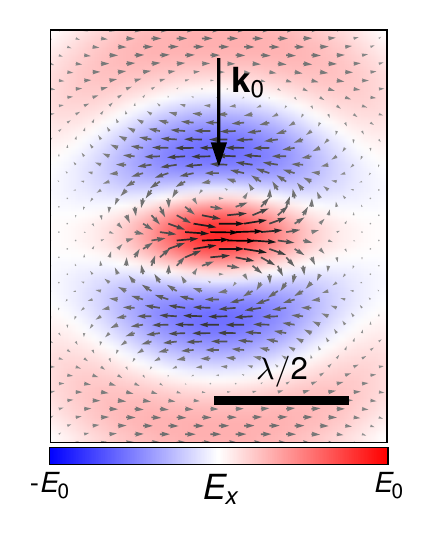}
  \includegraphics[width=.195\textwidth]{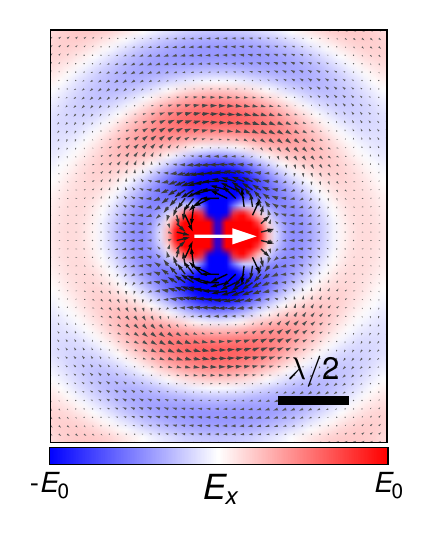}
  \caption{
  Real part of \(E_x\) for (from left to right): A plane wave, a ``focused plane wave'', a paraxial Gaussian beam, a tight-focus corrected paraxial Gaussian beam (all \(X\)-polarized, \(\mathbf{k} \parallel -\mathbf{e}_z\)) and a dipole emitter along \(X\) (indicated by a white arrow).}\label{fig:sim_fundamental_fields}
\end{figure*}

The information about the incident field is provided to \pygdm\ via
\functiondescription{o}{fields.efield}
  { {\textit{field\_generator:} field generator (for available functions, see below)} 
    {\textit{wavelengths:} \textbf{list} of wavelengths to calculate (nm)} 
    {\textit{kwargs:} \textbf{dict} with kwargs passed to the field generator. Can contain lists of multiple values for each parameter. All possible permutations will be calculated.} }
  {}
Illustrations of the below listed incident fields available in \pygdm\ are shown in figure~\ref{fig:sim_fundamental_fields}.

%

\subsubsection{Plane wave}

\functiondescription{f}{fields.planewave}
  { {\textit{theta:} linear polarization angle (degrees)} }
  {}

The probably most common fundamental field is the plane wave, which is in many cases a sufficient approximation. 
Its complex amplitude can be expressed as
\begin{equation}
 \mathbf{E}_0(\mathbf{r}, \omega) = \mathbf{E}_0\, \e^{\iu \mathbf{k}\cdot\mathbf{r}}.
 \label{eq:planeWaveField}
\end{equation}

\subsubsection{Focused plane wave}

\functiondescription{f}{fields.focused\_planewave}
  { {\textit{theta:} linear polarization angle (degrees)}
    {\textit{xSpot:} \(x\)-coordinate of focal spot (nm)} 
    {\textit{ySpot:} \(y\)-coordinate of focal spot (nm)} 
    {\textit{NA:} numerical aperture to calculate spotsize for each wavelength} 
    {\textit{spotsize:} Gaussian spotsize \(w\) in nm (required if \textit{NA}\,\(=-1\))} }
  {}

The simplest approximation for a focused beam is a plane wave with a Gaussian intensity profile. For incidence along \(Z\) (\(\mathbf{k} \parallel \mathbf{e}_z\)) this writes:
\begin{equation}
 \mathbf{E}_0(\mathbf{r}, \omega) = 
 \mathbf{E}_0\, \e^{\iu \mathbf{k}\cdot \mathbf{r}} 
 \exp \left( \dfrac{(x - x_0)^2 + (y - y_0)^2}{ 2 w_{\text{spot}}^2 } \right)
 \label{eq:focusedPlaneWaveField}
\end{equation}
The beam propagates along \((x_0,y_0,z)\). 
The full width at half maximum (FWHM) can be obtained via
\begin{equation}
 w_{\text{FWHM}} = w_{\text{spot}} \cdot 2\sqrt{2\ln 2} \, .
\end{equation}
A focused plane wave is often a sufficient approximation (see e.g. Ref.~\onlinecite{viarbitskaya_tailoring_2013}) and can be particularly useful if the divergence of the radius of curvature at the origin of the paraxial Gaussian becomes problematic.

\subsubsection{Paraxial Gaussian beam}

\functiondescription{f}{fields.gaussian}
  { {\textit{theta:} linear polarization angle (degrees)}
    {\textit{xSpot:} \(x\)-coordinate of focal spot (nm)} 
    {\textit{ySpot:} \(y\)-coordinate of focal spot (nm)} 
    {\textit{NA:} numerical aperture to calculate spotsize for each wavelength} 
    {\textit{spotsize:} Gaussian spotsize \(w\) in nm (required if \textit{NA}\,\(=-1\))} 
  }
  {{\textit{paraxial:} \textbf{True} (default: \textbf{False})}}

Often, lasers are used as sources of monochromatic, coherent light with high intensity. 
Light emitted from a laser-cavity is however not propagating like a plane wave, but as a Gaussian beam. 
The intensity profile differs significantly from the focused plane wave so the use of a model for Gaussian beams may become necessary -- particularly in larger objects, where the ``curved'' intensity profile of such a beam induces important field gradients along the propagation direction and the particle. 
A popular approximation to a real Gaussian beam is the so-called \emph{paraxial approximation}, where all \(\mathbf{k}\)-vectors are parallel to one single propagation direction. 
It can be calculated using the following formula (propagation along \(Z\)-axis)
\begin{multline}
 \mathbf{E}_0(\mathbf{r}, \omega) = 
  \mathbf{E}_0\, 
  \frac{w_0}{w(z)} \exp \left( \frac{-r^2}{w(z)^2} \right) \\
  \times \exp \left( -\iu \left( k\left(z + \frac{r^2}{2 R(z)}\right) - \zeta(z) \right) \right)
 \label{eq:paraxialGaussianField}
\end{multline}
with the beam width or ``waist'' \(w_0\) and the squared distance to the beam axis \(r^2 = (\Delta x^2 + \Delta y^2)\). 
\(\Delta x\) and \(\Delta y\) are the distances to the beam axis in \(X\) and \(Y\) direction, respectively.
In equation~\eqref{eq:paraxialGaussianField} we introduced furthermore the \(z\)-dependent beam waist
\begin{equation}
 w(z) = w_0\sqrt{1 + \left( \frac{z \lambda}{\pi w_0^2} \right)^2}
\end{equation}
the radius of curvature
\begin{equation}
 R(z) = z \left(1 + \left( \frac{\pi w_0^2}{z \lambda} \right)^2 \right)
\end{equation}
and the \emph{Gouy phase}~\cite{boyd_intuitive_1980}
\begin{equation}
 \zeta(z) = \arctan \left( \frac{z \lambda}{\pi w_0^2} \right).
\end{equation}

\subsubsection{Tightly focused Gaussian beam}

\functiondescription{f}{fields.gaussian}
  { {same as paraxial Gaussian.} }
  { {\textit{paraxial:} \textbf{False} (=default)} }

Under tight focusing conditions an additional component \(E_{0,z}\) parallel to the wave-vector (again assuming \(\mathbf{k} \parallel \mathbf{e}_z\)) can gain a substantial magnitude, which can be explained by the \(\text{div} \mathbf{E}\) Maxwell's equation.
This can be accounted for by adding the following correction term to the paraxial Gaussian (again assuming propagation along~\(Z\))~\cite{novotny_principles_2006}
\begin{equation}
 E_{0,z}(x,y,z) =  \frac{- 2 \iu}{k w(z)^2}  \left( \Delta x\, E_{0,x} + \Delta y\, E_{0,y} \right).
 \label{eq:gaussianCorrectField}
\end{equation}

\subsubsection{Dipolar emitter}

\functiondescription{f}{fields.dipole\_electric}
  { {\textit{x0, y0, z0:} position of dipole (in nm)}
    {\textit{mx, my, mz:} amplitude and direction of dipole-vector} }
  {}

An electric dipole \(\mathbf{p}\) placed in a homogeneous environment at \(\mathbf{r}_0\) and oscillating at frequency \(\omega\) creates an electric field at \(\mathbf{r}\) which writes~\cite{agarwal_quantum_1975}
\begin{equation}\label{eq:field_electric_dipolar_emitter}
 \mathbf{E}_{\text{p}} (\mathbf{r}, \mathbf{r}_0, \omega) = 
\frac{1}{\epsilon_{\text{env}}} \Big( \mathbf{I}\, k^2 + \mathbf{\nabla\nabla} \Big) G_{0} (\mathbf{r}, \mathbf{r}_0, \omega) \cdot \mathbf{p}(\omega)
\end{equation}
where \(\nabla\) acts along \(\mathbf{r}\), \(\mathbf{I}\) is the unitary tensor, \(k\) the wavenumber and \(G_{0}\) the scalar vacuum Green's function (see Eq.~\eqref{eq:scalarGreenFunctionVac}).

\subsubsection{Magnetic dipole emitter}

\functiondescription{f}{fields.dipole\_magnetic}
  { {\textit{x0, y0, z0:} position of dipole (in nm)}
    {\textit{mx, my, mz:} amplitude and direction of dipole-vector} }
  {}

Analogously, a magnetic dipole emitter \(\mathbf{m}\) at \(\mathbf{r}_0\) is the source of an electric field~\cite{agarwal_quantum_1975, wiecha_decay_2018}
\begin{equation}\label{eq:field_magnetic_dipolar_emitter}
 \mathbf{E}_{\text{m}} (\mathbf{r}, \mathbf{r}_0, \omega) = 
 \iu k_0 \nabla \times G_{0} (\mathbf{r}, \mathbf{r}_0, \omega) \cdot \mathbf{m}(\omega).
\end{equation}

\section{Solver}\label{sec:solver}

\subsection{Internal fields}

Solving the primary scattering problem is usually the root of a GDM simulation. 
The self-consistent calculation of the fully retarded (complex) electric field inside the nano-structure is done by inversion of equation~\eqref{eq:definitionMainEquationGDM} via

\functiondescription{f}{core.scatter}
  { {\textit{sim}: instance of \object{core.simulation}} }
  {}
Usually, the underlying \textit{scipy} libraries used for inversion are multi-thread parallelized, making use of all processors on multi-core CPUs. 

For distributed systems like most modern computing clusters, also a multi-processing parallel version of \function{core.scatter} is implemented in \pygdm , which uses MPI to simultaneously calculate several wavelengths of a spectrum on parallel processes:
\functiondescription{f}{core.scatter\_mpi}
  { {\textit{sim}: instance of \object{core.simulation}} }
  {}
\paragraph*{Note:}
Each MPI process calculates a single wavelength using the parallelized \textit{scipy}-routines. In this double-parallelized way, spectral simulations can be carried out very rapidly on multi-node computing clusters.
\function{core.scatter\_mpi} requires the ``mpi4py'' package.

%
\begin{figure*}[t]
%
  \centering
%
%
  \includegraphics{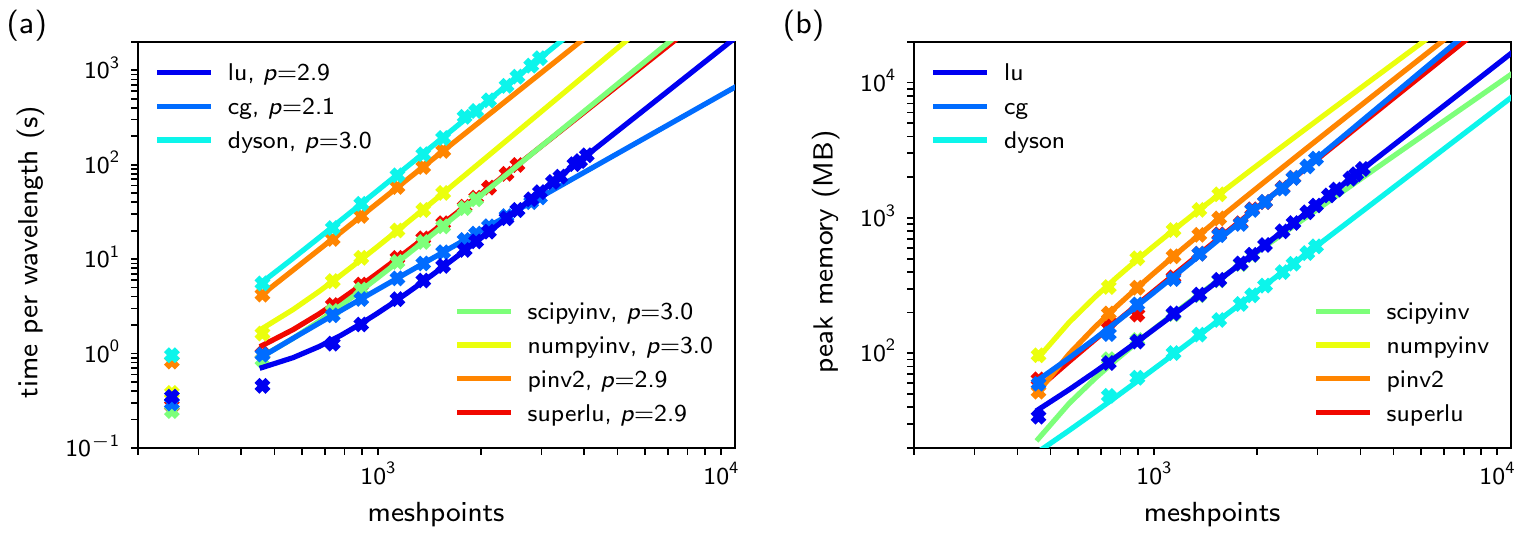}
  \caption{
  %
  (a) Timings of a \pygdm-simulation of a spherical dielectric particle as a function of the number of meshpoints for the different available solvers. 
  Solid lines are power-law fits, confirming \(p=3\) for full inversion methods and \(p=2\) for CG (the fitted power \(p\) is given in the legend).
  (b) Memory requirement (in megabytes) as function of the number of meshpoints for the different solvers.
  All benchmarks were performed on a single core of an AMD FX-8350 CPU.
  }\label{fig:Theory_inversion_in_GDM}
\end{figure*}

%
\begin{figure}[t]
  \centering
  \includegraphics{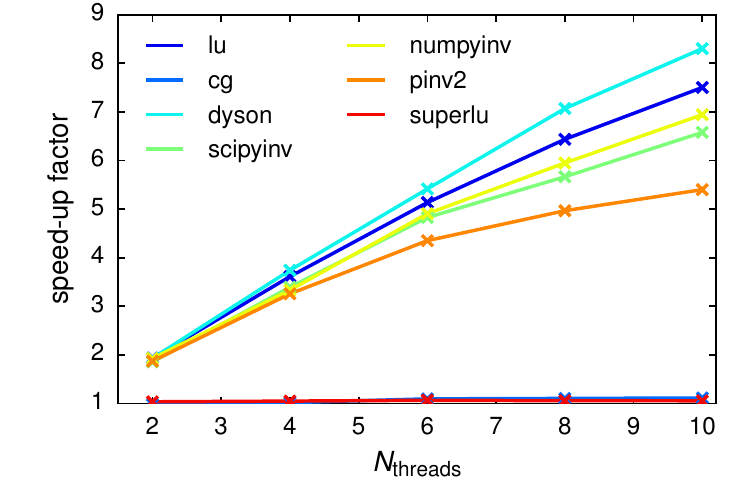}
  \caption{
  Speedup of the GDM-calculation using the multi-threaded parallelization capability of the available solvers. 
  Benchmark performed on an Intel E5-2680 10-core CPU.}\label{fig:parallelization_speedup}
\end{figure}
%

\subsubsection{Direct inversion}

\begin{itemize}
 \item argument \textit{method}: ``lu'' (default), ``scipyinv'', ``superlu'', ``pinv2'' (all require \textit{scipy}), ``numpyinv'' or ``dyson'' (only \textit{numpy})
\end{itemize}

In \pygdm\ the inversion of \(\mathbf{M}\) (eq.~\eqref{eq:definitionMforInversion}) is by default performed with LU-decomposition (using the implementation in \textit{scipy}). 
This should be the fastest solver for full inversion (see Fig.~\ref{fig:Theory_inversion_in_GDM}a) and has furthermore an excellent multi-threaded parallelization scaling, as can be seen in figure~\ref{fig:parallelization_speedup}.
An extensive explanation of LU-decomposition and details on its implementation can be found for example in Ref.~\onlinecite{press_numerical_2007} (chapter~2.3). 
Other scipy solvers can be used in \pygdm, and, if for any reason \textit{scipy} is not available, the ``numpyinv'' and ``dyson'' methods are alternatives which do not require \textit{scipy}.

The solver ``dyson'' uses a sequence of Dyson's equations~\cite{martin_generalized_1995} and comes with \pygdm. 
Since it does not depend on any libraries it should work in every case, however it will usually be significantly slower than the third-party solvers. 
An advantage of ``dyson'' can be the memory requirement which is relatively low, because the Dyson sequence allows an in-place inversion of the matrix (see figure~\ref{fig:Theory_inversion_in_GDM}b).
A detailed description of the latter algorithm can be found in Ref.~\onlinecite{colas_des_francs_optique_2002} (chapter~2.4). 

We note that LU inversion (or in some cases conjugate gradients, e.g. for dense spectra on single-core systems, see below and appendix) is the preferred technique in \pygdm\ due to its high efficiency (see Fig.~\ref{fig:Theory_inversion_in_GDM}a).

\subsubsection{Conjugate gradients}

\begin{itemize}
 \item argument \textit{method}: ``cg'' or ``pycg''
\end{itemize}

Sometimes it is not necessary to calculate the full structure of the inverse of matrix \(\mathbf{M}\). 
Often it is sufficient to only know the \textit{result} of the matrix-vector product \(\mathbf{M}^{-1} \mathbf{E}_0\). It turns out that under certain circumstances, iterative approaches such as the "conjugate gradients" method lead to very accurate approximations of this matrix/vector product in significantly less time compared to the inversion of \(\mathbf{M}\).
For a detailed description and informations related to the conjugate gradients solver, see appendix~\ref{sec:conjugate_gradients}.

\subsection{Decay-rate of dipolar emitters}\label{sec:e_m_decay}

\functiondescription{f}{core.decay\_rate}{}{}

The Green's Dyadic formalism can be used not only to obtain scattered electro-magnetic fields. 
It gives also direct access to the modification of the decay rate of electric or magnetic dipolar transitions due to the presence of polarizable materials in their vicinity. 

\paragraph*{Note: } The decay rates are proportional to the photonic LDOS~\cite{carminati_electromagnetic_2015}, hence the values obtained from the calculation of the relative decay rates \(\Gamma / \Gamma_0\) are identical to the relative LDOS (specifically to the \textit{partial} LDOS, meaning its electric or magnetic component and / or partial for specific dipole orientations).

\subsubsection{Electric dipole}\label{sec:e_decay}

The effect is intuitively understandable for an electric dipole transition \(\mathbf{p}\), as a consequence of the enhancement (or weakening) of the electric near-field because of the dielectric contrast and the resulting back-action of the radiated field on the dipole itself.
It is possible to derive an integral equation describing the decay rate \(\Gamma_{e}\) of the dipole transition~\cite{carminati_electromagnetic_2015, wiecha_decay_2018}:
\begin{multline}\label{eq:gamma_electric}
\Gamma_{e}({\bf r}_{0},\omega) = \Gamma_{e}^{0}(\omega)
\\ 
\times \left(1+
\frac{3}{2k_{0}^{3}}{\bf u}\cdot \text{Im}\big(
{\boldsymbol{\cal G}}_{p}^\text{EE}({\bf r}_{0},{\bf r}_{0},\omega)\big)\cdot{\bf u}
\right) \; ,
\end{multline} 
where
\begin{multline}\label{eq:gamma_magnetic_SpEE}
{\boldsymbol{\cal G}}_{p}^\text{EE}({\bf r},{\bf r}_{0},\omega) =
\int\limits_{V}\text{d}\mathbf{r}'\int\limits_{V}\text{d}\mathbf{r}''\mathbf{G}_{0}^{\text{EE}}({\bf r},{\bf r}',\omega)
\\
\cdot \boldsymbol\chi_{\text{e}}({\bf r}',\omega) 
\cdot \mathbf{K}({\bf r}',{\bf r}'',\omega)\cdot\mathbf{G}_{0}^\text{EE}({\bf r}'',{\bf r}_{0},\omega)
\; 
\end{multline}
and \(\Gamma_{e}^{0}(\omega)$ = $ 4k_{0}^{3} p^{2}/3\hbar\) is the decay rate of the electric dipole transition in vacuum. 
$\mathbf{r}_0$ is the location of the dipolar transition (outside the nanostructure). ${\bf u}$ denotes the dipole orientation, \(p\) its amplitude. 
\(\mathbf{K}\) is the generalized propagator (see Eq.~\eqref{eq:definitionGeneralizedPropoagator} or Ref.~\onlinecite{martin_generalized_1995}).
For the numerical implementation, the integrals in Eq.~\eqref{eq:gamma_magnetic_SpEE} become sums over the mesh-points of the discretized nano-object(s).

The propagator \(\mathbf{G}_{0}^\text{EE}\) can be found by identification using Eq.~\eqref{eq:field_electric_dipolar_emitter} and following equation for the field of an electric dipole \(\mathbf{p}\) at \(\mathbf{r}_0\)~\cite{agarwal_quantum_1975}
\begin{equation}\label{eq:gamma_magnetic_SEE}
{\bf E}_{0}({\bf r},\omega) 
= \mathbf{G}_{0}^\text{EE}({\bf r},{\bf r}_{0},\omega)\cdot{\bf p}(\omega) \; ,
\end{equation}
\paragraph*{Note:} \(\mathbf{G}_{0}^\text{EE}\) is given for particles in a homogeneous environment by the Green's Dyad of equation~\eqref{eq:vacuumGreenDyadicFunction}. 
In analogy to the scattering simulations it can be easily extended for more complex environments, such as an infinite substrate (see section~\ref{sec:green_method}).

\subsubsection{Magnetic dipole}\label{sec:m_decay}

Also the decay rate of a \textit{magnetic} dipole transition close to \textit{non-magnetic} materials is influenced by the presence of the structure.
Such magnetic-magnetic response function associated to a structure with no direct magnetic response, arises from the \textit{electric} field emitted by the magnetic dipole, interacting with the material and finally again inducing a magnetic field via the curl of the electric field.
In metallic nanostructures, circular plasmonic currents can also lead to significant magnetic near-field enhancements~\cite{huidobro_magnetic_2014}.
In complete analogy to equation~\eqref{eq:gamma_electric}, the magnetic decay rate \(\Gamma_{m}\) writes
\begin{multline}\label{eq:gamma_magnetic}
\Gamma_{m}({\bf r}_{0},\omega)=\Gamma_{m}^{0}(\omega)
\\ 
\times \left(1+
\frac{3}{2k_{0}^{3}}{\bf u}\cdot \text{Im}\big(
{\boldsymbol{\cal G}}_{p}^\text{HH}({\bf r}_{0},{\bf r}_{0},\omega)\big)\cdot{\bf u}
\right) \; ,
\end{multline} 
where
\begin{multline}\label{eq:gamma_magnetic_SpHH}
{\boldsymbol{\cal G}}_{p}^\text{HH}(\mathbf{r},\mathbf{r}_{0},\omega) =
\int\limits_{V}\text{d}\mathbf{r}'\int\limits_{V}\text{d}\mathbf{r}'' \mathbf{G}_{0}^\text{HE}({\bf r},{\bf r}',\omega)
\\
\cdot \boldsymbol\chi_{\text{e}}({\bf r}',\omega) 
\cdot\, \mathbf{K}({\bf r}',{\bf r}'',\omega)\cdot\mathbf{G}_{0}^\text{EH}({\bf r}'',{\bf r}_{0},\omega)
\; 
\end{multline}
and \(\Gamma_{m}^{0}(\omega)$ = $ 4k_{0}^{3} m^{2}/3\hbar\) is the decay rate of the magnetic transition in vacuum. 
${\bf u}$ and \(m\) are the magnetic dipole orientation and amplitude, respectively, and \(\mathbf{K}\) is again the generalized propagator.

In the same way as for the electric dipole, \(\mathbf{G}_{0}^\text{HE}\) and \(\mathbf{G}_{0}^\text{EH}\) can be found using Eq.~\eqref{eq:field_magnetic_dipolar_emitter} with the electric field of a magnetic dipole \(\mathbf{m}\) at \(\mathbf{r}_0\)
\begin{equation}\label{eq:gamma_magnetic_SEH}
{\bf E}_{0}({\bf r},\omega) 
= \mathbf{G}_{0}^\text{EH}({\bf r},{\bf r}_{0},\omega)\cdot{\bf m}(\omega)
\end{equation}
and 
\begin{equation}\label{eq:gamma_magnetic_Efield_m}
\mathbf{G}_{0}^\text{HE}({\bf r},{\bf r}',\omega)
= \mathbf{G}_{0}^\text{EH}({\bf r}',{\bf r},\omega) \; .
\end{equation}
For a detailed derivation of the formalism see reference~\onlinecite{wiecha_decay_2018}. For a comparison of our code with experimental results, see reference~\onlinecite{wiecha_simultaneous_2018}.

\subsubsection{LDOS inside a nanostructure}\label{sec:decay_inside}

The decay rate (and hence the LDOS) at a position ${\bf r}_{0,s}$ \textit{inside} the structure can also be obtained via equation~\eqref{eq:gamma_electric} (for the electric case), using the field susceptibility \({\boldsymbol{\cal G}}_{p,s}^\text{EE}\) inside the structure. 
It is related to the generalized propagator (assuming an isotropic medium with \(\chi_{\text{e}} = \text{Tr}\,(\boldsymbol\chi_{\text{e}})/3\)), by
\begin{equation}
{\boldsymbol{\cal G}}_{p,s}^\text{EE}({\bf r}_i,{\bf r}_j,\omega) = 
\frac{\mathbf{K}({\bf r}_i,{\bf r}_j,\omega) - \mathbf{I}}{\chi_{\text{e}}\, V_{\text{cell}}} \, ,
\end{equation}
where \(\mathbf{r}_i\) and \(\mathbf{r}_j\) are positions of nano-particle meshpoints.

For the case of the magnetic LDOS inside the structure, an ``electric-magnetic'' mixed generalized propagator needs to be calculated. 
This propagator relates any incident \textit{electric} field to the scattered \textit{magnetic} field inside the structure.
This is not implemented in \pygdm\ so far, but could easily be made available using the mixed tensor \(\mathbf{G}_{0}^{\text{EH}}\) instead of the electric-electric Green's Dyadic function in the inversion problem, defined by equation~\eqref{eq:LippmannSchwingerVolumeDiscretization}.

\paragraph*{Note:} The frequency shift of the emitter due to the presence of a nano-structure (``Lamb shift'') can be obtained in analogy to the decay rate, via the real part of the field susceptibility \cite{lassalle_lamb_2017}. 
This is however not (yet) implemented in \pygdm.

\section{Post-Processing}\label{sec:postprocessing}

\subsection{Linear effects}

After the main simulation (calculation of the fields inside the structure, decay rate), the information can be further processed to obtain experimentally accessible physical quantities.

\subsubsection{Near-field outside the nanostructure}\label{sec:Theory_NF_outside_structure}

\functiondescription{f}{linear.\allowbreak nearfield}
  { {\textit{sim}: instance of \object{core.simulation}} 
    {\textit{field\_index}: index of field-configuration (see section~\ref{sec:tools_get_field_index}: \function{tools.\allowbreak get\_\allowbreak closest\_\allowbreak field\_\allowbreak index})} 
    {\textit{MAP}: \textbf{list} of \(X\), \(Y\) and \(Z\) coordinates at which to evaluate the near-field} }
  {}

\paragraph*{Electric field:}
Via Eq.~\eqref{eq:LippmannSchwingerVolumeDiscretization} the electric field induced at any point \(\mathbf{r}\) at the exterior of the particle can be calculated from the electric polarization inside the structure.

\paragraph*{Magnetic field:}

The propagator \(\mathbf{G}_{0}^{\text{HE}}\) (see also equation~\eqref{eq:gamma_magnetic_SEH}) can be used to obtain the magnetic field outside the source region~\cite{girard_optical_1997}.
\function{linear.\allowbreak nearfield} returns both, the electric and the magnetic field amplitudes for the scattered as well as for the total near-field (\(\mathbf{E}_{\text{tot}} = \mathbf{E}_{\text{scat}} + \mathbf{E}_0\)).

\paragraph*{Note:} 
Alternatively, the \(\mathbf{B}\)-field may be calculated via finite differentiation: 
After Faraday's induction law from Maxwell's equations (Eq.~\eqref{eq:MaxwellFourierrotE}), the magnetic field writes (for time-harmonic fields)
\begin{equation}
  \mathbf{B}(\mathbf{r}, \omega) 
=
  \frac{\nabla \times \mathbf{E}(\mathbf{r}, \omega) }{\iu k_0}.
\end{equation}

\subsubsection{Extinction, absorption and scattering cross-sections}\label{sec:Theory_Spectra_from_NF}

\functiondescription{f}{linear.\allowbreak extinct}
  { {\textit{sim}: instance of \object{core.simulation}} }
  {}

The linear response in the farfield can be characterized by the scattered and absorbed light intensity, the sum of which is called the ``extinction''.
Usually these values are given as cross sections \(\sigma_{\text{scat.}}, \sigma_{\text{abs.}}\) and \(\sigma_{\text{ext.}}\) which have the unit of an area.
The extinction and absorption cross sections can be calculated from the near-field in the discretized structure~\cite{draine_discrete-dipole_1988}
\begin{equation}\label{eq:sigma_ext_from_nearfield}
 \sigma_{\text{ext}} = \frac{4 \pi k}{|E_0|^2} 
	    \sum\limits_{i=1}^{N_\text{cells}}\  
	\text{Im} \left( \mathbf{E}_{0,i}^* \cdot \mathbf{P}_i \right)
\end{equation}
and
\begin{equation}\label{eq:sigma_abs_from_nearfield}
 \sigma_{\text{abs}} = \frac{4 \pi k}{|E_0|^2} 
	    \sum\limits_{i=1}^{N_\text{cells}}\ 
	\left( \text{Im} \left( \mathbf{P}_i \cdot \mathbf{E}_i^* \right) 
	        - \frac{2}{3} k^3 |\mathbf{P}_i|^2 \right).
\end{equation}
\(\mathbf{E}_i\) and \(\mathbf{P}_i\) are the electric field and polarization at meshpoint \(i\), respectively, induced by an excitation field \(\mathbf{E}_{0,i}\). 
\(k\) is the wavenumber in the particle's environment.
Complex conjugation is indicated with a superscript asterisk~(\(^*\)).

The scattering cross section finally is the difference of extinction and absorption
\begin{equation}\label{eq:sigma_scat_from_nearfield}
 \sigma_{\text{scat}} = \sigma_{\text{ext}} - \sigma_{\text{abs}}.
\end{equation}

\subsubsection{Far-field pattern of the scattered light}\label{sec:linear_farfield}

\functiondescription{f}{linear.\allowbreak farfield}
  { {\textit{sim}: instance of \object{core.simulation}} 
    {\textit{field\_index}: index of field-configuration (see section~\ref{sec:tools_get_field_index}: \function{tools.\allowbreak get\_\allowbreak closest\_\allowbreak field\_\allowbreak index})} }
  {}

The complex electric field in the far-field, radiated from an arbitrary polarization distribution can be calculated using a corresponding Green's Dyad \( \mathbf{G}_{\text{ff}}\) (assuming a dipolar emission from each of the \(N\) meshpoints):
\begin{equation}\label{eq:FarfieldFromNearfield}
 \mathbf{E}_{\text{ff}}(\mathbf{r}) = \sum\limits_i^{N_{\text{cells}}} \mathbf{G}_{\text{ff}}(\mathbf{r}_i, \mathbf{r}) \cdot \mathbf{P}(\mathbf{r}_i).
\end{equation}
In vacuum, using equation~\eqref{eq:vacuumGreenDyadicFunction} with only the far-field term \(\mathbf{T}_1\), we can calculate the electric field at any point \(\mathbf{r}\) far enough from the scatterer.

A substrate can be included in the asymptotic tensor by means of an appropriate dyadic Green's function. 
An analytic approximation of a farfield-propagator for a layered system has been derived e.g. by Novotny~\cite{novotny_allowed_1997}.
Making use of the superposition principle, the radiation of single dipoles via the propagator \(\mathbf{G}_{\text{ff}}\) can be generalized to the total far-field radiation of an ensemble of \(N\) dipole-emitters by simple summation of all meshpoints' contributions (see Eq.~\eqref{eq:FarfieldFromNearfield}).

Note that the presence of the illuminated nano-structure is fully taken into account also in this scattering formalism, thanks to the self-consistent nature of the Green's method.

Particularly in nano-structures with high absorption, equation~\eqref{eq:sigma_scat_from_nearfield} requires a high accuracy of the extinction and absorption cross-sections, hence small discretization steps, which can be practically not feasible~\cite{draine_discrete-dipole_1988}.
In such case, equation~\eqref{eq:FarfieldFromNearfield} offers a more precise alternative to determine the scattering cross-section.
A further drawback of the calculation of the scattering spectra from the near-field via Eq.~\eqref{eq:sigma_scat_from_nearfield} is obvious: 
These spectra do not contain any information about the directionality of the scattering.
Using Eq.~\eqref{eq:FarfieldFromNearfield} on the other hand, the spatial distribution and polarization of scattered light in the far-field and can be obtained.

In \pygdm, \function{linear.farfield} implements a Green's dyad including the contribution of an optional dielectric substrate (in a non-retarded approximation~\cite{novotny_allowed_1997}).

\subsubsection{Heat generation}

Having calculated the electric fields inside a nano-object, it is possible to compute the heat deposited inside the nanoparticle by an optical excitation as well as the temperature rise in the vicinity of the structure~\cite{baffou_heat_2009}.

\functiondescription{f}{linear.\allowbreak heat}
  { {\textit{sim}: instance of \object{core.simulation}} }
  {}
The total heat generated inside the nanoparticle is the product of the imaginary part of the material's permittivity and the electric field intensity:
\begin{equation}\label{eq:heat_deposited}
\begin{aligned}
 Q(\omega) =\ &
 \int\limits_V q(\mathbf{r}, \omega)\, \text{d}\mathbf{r} \\
  =\ &
 \frac{\omega}{8\pi} \int\limits_V  \mathrm{Im}\big(\epsilon (\mathbf{r})\big) \left| \mathbf{E}(\mathbf{r}, \omega) \right|^2 \text{d}\mathbf{r}.
\end{aligned}
\end{equation}
\functiondescription{f}{linear.\allowbreak temperature}
  { {\textit{sim}: instance of \object{core.simulation}} }
  {}
The temperature rise at position \(\mathbf{r}_{\text{probe}}\) outside the nanoparticle can be approximated with the heat \(q(\mathbf{r}, \omega)\), generated at each meshpoint (located at \(\mathbf{r}\)) via the thermal Poisson's equation~\cite{baffou_thermoplasmonics_2010, teulle_scanning_2012}
\begin{equation}\label{eq:temp_rise_vicinity}
\begin{aligned}
 \Delta T (\mathbf{r}_{\text{probe}}, \omega) = & \frac{1}{4\pi \kappa_{\text{env}}} \int\limits_V 
 \Bigg( 
 \frac{ q(\mathbf{r}, \omega) }{\left| \mathbf{r}_{\text{probe}} - \mathbf{r} \right|}  \\
 &  + \bigg(\frac{\kappa_{\text{sub}} - \kappa_{\text{env}}}{\kappa_{\text{sub}} + \kappa_{\text{env}}}\bigg) 
 \frac{ q(\mathbf{r}, \omega) }{\left| \mathbf{r}_{\text{probe}} - \mathbf{r} \right|}
 \Bigg)\text{d}\mathbf{r}
\end{aligned}
\end{equation}
where \(\kappa_{\text{env}}\) and \(\kappa_{\text{sub}}\) are the heat conductivities of the environment and substrate, respectively. 
The second term in the integrand can be derived through a formalism similar to image charges in electro-dynamics and accounts for heat reflection at the interface of the substrate~\cite{baffou_thermoplasmonics_2010}.
Eqs.~\eqref{eq:heat_deposited} and~\eqref{eq:temp_rise_vicinity} can be for example used to calculate raster-scan mappings of the deposited heat or the temperature increase as function of a focused beam's focal spot position.
Eq.~\eqref{eq:temp_rise_vicinity} can also be used to compute maps of the temperature increase above a nanostructure, by raster-scanning \(\mathbf{r}_{\text{probe}}\) under constant illumination conditions.

\paragraph*{Note:} 
Equation~\eqref{eq:temp_rise_vicinity} assumes that the heat \(q\) generated by the optical excitation at each meshpoint induces a static heat distribution inside the nanoparticle.
This approximation might become inaccurate in large nanoparticles of material with high heat conductivity (e.g. metals), leading to a rapid redistribution of the heat inside the nanostructure~\cite{baffou_heat_2009}.
If the temperature increase is evaluated at sufficiently large distances to the nano-object, Eq.~\eqref{eq:temp_rise_vicinity} is usually a good approximation also for larger metallic nano-objects~\cite{wiecha_local_2017}.
``Sufficiently large distances'' could mean comparable to, or larger than the size of the nanoparticle.

\subsubsection{Dipolar emitter decay rate}

\functiondescription{f}{linear.decay\_eval}{}{}

The decay rate of magnetic or electric dipole emitters can be calculated within the GDM as described in section~\ref{sec:e_m_decay}.
Via equation~\eqref{eq:gamma_electric}, the tensor \({\cal S}_{p}^\text{EE}\) (or  \({\cal S}_{p}^\text{HH}\) using Eq.~\eqref{eq:gamma_magnetic}) can be used to calculate the decay rate of the transition for arbitrary orientations of the dipole.

\function{core.decay\_\allowbreak rate} calculates the tensor \({\cal S}_{p}^\text{EE}\) or \({\cal S}_{p}^\text{HH}\) (for an electric, respectively magnetic dipole emitter) at each user-defined dipole position and wavelength. 
The final evaluation of the decay rate is done using \function{linear.decay\_eval} for a given dipole orientation and amplitude.
The advantage of this two-step approach is that the generalized propagator needs to be computed only once, and the results of this expensive part of the simulation can be re-used for multiple dipole orientations and/or amplitudes.

\subsection{Non-linear effects}\label{sec:nonlinear_effects}

\subsubsection{Two-photon photoluminescence / surface LDOS}

\functiondescription{f}{nonlinear.tpl\_ldos}{}{}
Having calculated the electric field distribution inside a nanoparticle, a simple model allows to calculate the two-photon photoluminescence (TPL) signal   generated by the excitation: 
We assume that the TPL is proportional to the square of the electric field intensity.
We furthermore consider each meshpoint (at position \(\mathbf{r}\)) as an incoherent source of TPL, contributing to the total TPL with an intensity proportional to \(|\mathbf{E}(\mathbf{r}, \omega)|^4\). 
Integration over the nano-particle volume \(V\) results in the total TPL intensity~\cite{teulle_scanning_2012}:
 \begin{equation}\label{eq:Intensity_TPL}
  I_{\text{TPL}}(\mathbf{r}_{\text{focus}}, \omega) \propto 
  \int\limits_{V} \left| \mathbf{E}(\mathbf{r}, \mathbf{r}_{\text{focus}}, \omega) \right|^4 \text{d}\mathbf{r}.
 \end{equation}
Here we added a further parameter, the focal spot position \(\mathbf{r}_{\text{focus}}\) of a focused illumination. 
By performing a raster-scan over the nano-structure with the focal position coordinate, we can calculate \(2\)D scanning TPL-maps.

This approach allows also to approximate the photonic local density of states at the surface of the nanostructure \(\rho_{\text{sf}}(\mathbf{r}, \omega)\) on which the focused spot impinges (surface LDOS), using an unphysically tightly focused beam. 
In the case of a circularly polarized excitaiton, it is possible to rewrite the TPL intensity of Eq.~\eqref{eq:Intensity_TPL}~\cite{teulle_scanning_2012, viarbitskaya_tailoring_2013, viarbitskaya_plasmonic_2015}:
 \begin{multline}\label{eq:Intensity_TPL_LDOS}
  I_{\text{TPL}}(\mathbf{r}_{\text{focus}}, \omega) \propto  \\
  \int\limits_{V} \left| \mathbf{E}_0^{\rcirclearrow}(\mathbf{r}, \mathbf{r}_{\text{focus}}, \omega) \right|^4 \rho_{\text{sf}, \parallel}^2(\mathbf{r}, \omega) \text{d}\mathbf{r}
 \end{multline}
where \(\mathbf{E}_0^{\rcirclearrow}\) is the incident electric field and \(\rho_{\text{sf}, \parallel}\) is the component of the LDOS in the plane parallel to the incident electric field vector.
Let us now decrease the waist of the focused beam: In the limit of a spatial profile of \(\mathbf{E}_0^{\rcirclearrow}\) corresponding to a Dirac delta function, the square root of the TPL intensity Eq.~\eqref{eq:Intensity_TPL_LDOS} becomes proportional to the LDOS at the position of the focal spot
 \begin{equation}\label{eq:Intensity_TPL_propto_LDOS}
  I_{\text{TPL}}(\mathbf{r}_{\text{focus}}, \omega) \propto  \\
  \rho_{\text{sf}, \parallel}^2(\mathbf{r}_{\text{focus}}, \omega).
 \end{equation}
In consequence, a \(2\)D map of the LDOS can be calculated via a raster-scan simulation, which can be done very efficiently in \pygdm\ thanks to the generalized propagator.
By using a linear polarized incident field, it is furthermore possible to extract partial contributions to the LDOS for the corresponding polarization.

\paragraph*{Note:}
The ``surface''-LDOS is reproduced by Eq.~\eqref{eq:Intensity_TPL_propto_LDOS} for a contraction of \(\mathbf{E}_0^{\rcirclearrow}\) towards a Dirac delta function.
However, due to the finite stepsize in the GDM, the beam waist cannot be reduced to an infinitely small value, hence this method remains approximative.
Practical values for the waist must be at least as large as a few times the discretization stepsize.
To obtain the exact LDOS, the calculation of the decay-rate is the method of choice (see also section~\ref{sec:e_m_decay}).

%
%
%

\section{Visualization}\label{sec:visu}

\pygdm\ includes several visualization tools for simple and rapid plotting of the simulation results.
They are divided into functions for the visualization of \(2\)D representations and functions for \(3\)D plots.

\subsection{2D visualization tools}

%
\begin{figure*}[t]
  \centering
  \includegraphics[width=\textwidth]{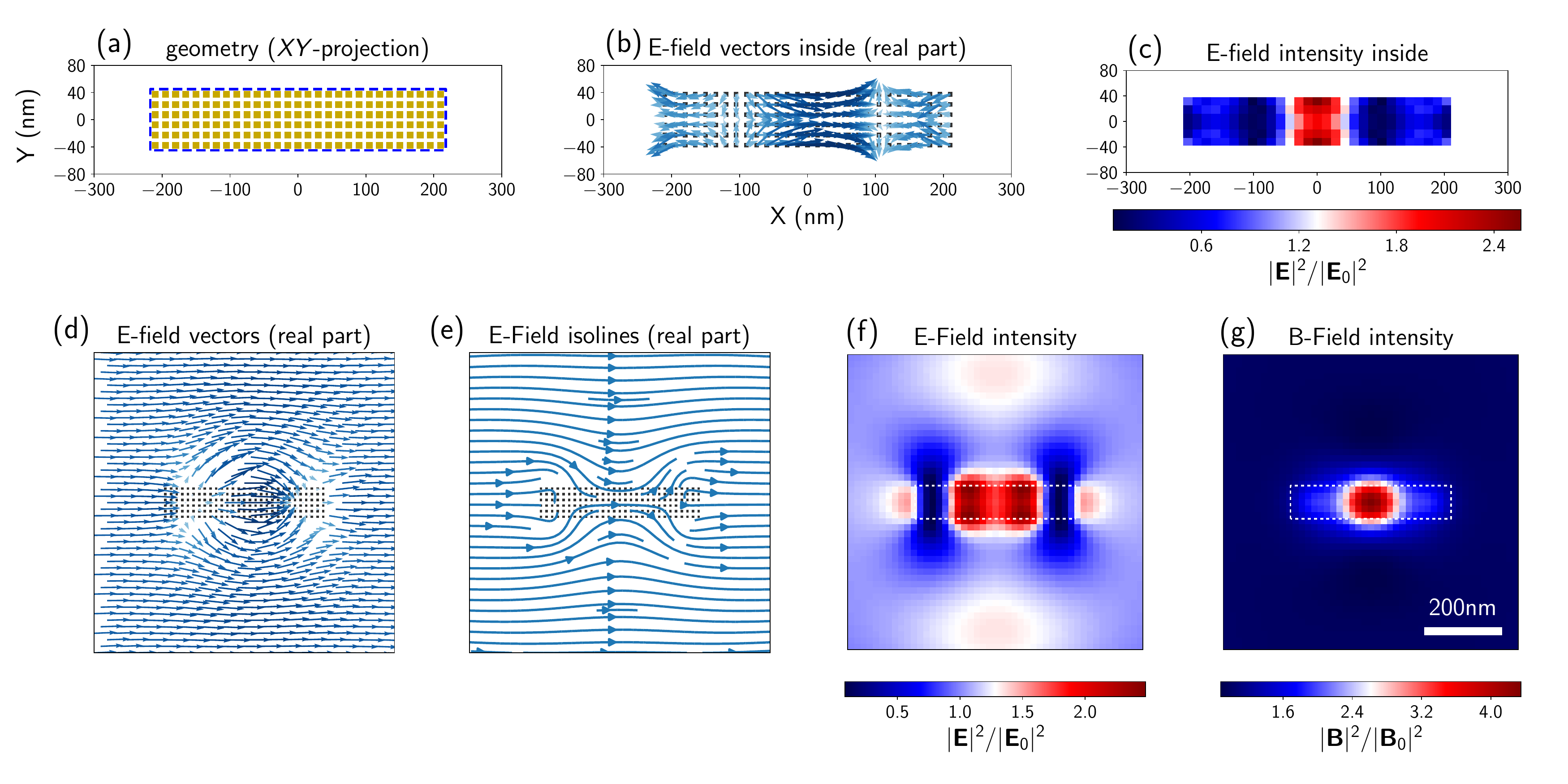}
  \caption{
  Visualization tools available in \pygdm\ on the example of a \(450\times 90\times 45\,\)nm\(^3\) (\(L\times W\times H\)) gold-rod placed in vacuum.
  Plane wave illumination incident along \(-Z\), linear polarization along \(X\), \(\lambda=600\,\)nm. 
  All plots show projections on the \(XY\)-plane.
  (a) the geometry (gold) and its surface contour (dashed blue), (b) the real part of the internal electric field and (c) the internal electric field intensity at the bottom of the rod.
  (d-g) show external fields, calculated on an \(800\times 800\,\)nm\(^2\) large area, \(30\,\)nm below the structure using \function{linear.nearfield}: (d) \textbf{E}-field real part, (e) isolines of \textbf{E}-field, (f) electric field intensity and (g) magnetic field intensity.
  }\label{fig:tools_visu}
\end{figure*}
%

The available visualization functions are explained in the following, examples are given in Fig.~\ref{fig:tools_visu} using a simulation of a \(450\,\)nm \(\times\ 90\,\)nm large gold rod with stepsize \(\stepsize=15\,\)nm, excited with a plane wave at \(\lambda_0 = 600\,\)nm, linearly polarized along \(X\) and incident from the reader towards the paper (\(\mathbf{k} = -\mathbf{e}_z \, k\)).
The plots show projections on the \(XY\) plane.

\subsubsection{Structure geometry}

\functiondescription{f}{visu.structure}
  { {\textit{sim}: instance of \object{core.simulation}} } 
  {}

Plot a \(2\)D projection of the simulated nano-particle geometry (see figure~\ref{fig:tools_visu}a, meshpoints in golden color).

\functiondescription{f}{visu.structure\_\allowbreak contour}
  { {\textit{sim}: instance of \object{core.simulation}} } 
  {}

Plot a contour around a \(2\)D projection of the nano-particle, in other words drawing the outer surface of the structure (see figure~\ref{fig:tools_visu}, dashed blue line in (a), dashed white lines in (f-g)).

\subsubsection{Plot field vectors}

\(2\)D projections of the real or imaginary part of vector-fields (see figure~\ref{fig:tools_visu}d) can be plotted using
\functiondescription{f}{visu.vectorfield}
  { {\textit{NF}: list containing the complex field (list of 6-tuples \([x_i, y_i, z_i, E_{x,i}, E_{y,i}, E_{z,i}]\), see section~\ref{sec:tools_get_field_as_list}: \function{tools.\allowbreak get\_\allowbreak field\_\allowbreak as\_\allowbreak list})} }
  {}
Alternatively, the function can be called using:
\functiondescription{f}{visu.vectorfield\_\allowbreak by\_\allowbreak fieldindex}
  { {\textit{sim}: instance of \object{core.simulation}} 
    {\textit{field\_index}: index of field-configuration (see section~\ref{sec:tools_get_field_index}: \function{tools.\allowbreak get\_\allowbreak closest\_\allowbreak field\_\allowbreak index})} }
  {}
The intention of the latter is to be used for direct plotting of fields inside the simulated particle via the \object{core.simulation} object (see figure~\ref{fig:tools_visu}b).

\subsubsection{Field lines (``stream-plot'')}

Isolines of the field amplitude can be plotted using
\functiondescription{f}{visu.vectorfield\_\allowbreak fieldlines}
  { {\textit{NF}: list containing the complex field} }
  {}
For an example, see figure~\ref{fig:tools_visu}e.

\subsubsection{Scalar field representation (color-plot)}

Color-plots are well suited to illustrate a scalar representation of the electric- or magnetic-field. 
This can be used to represent either the real/imaginary part of an individual field component (such as \(E_x\)), or the field intensity (\(|\mathbf{E}|^2\), \(|\mathbf{B}|^2\)).
In \pygdm, such a plot can be drawn using
\functiondescription{f}{visu.vectorfield\_\allowbreak color}{}{}
By default, the electric field intensity is plotted, as shown in figure~\ref{fig:tools_visu}f-g. 
Alternatively, to easily plot the field inside the nanoparticle (see figure~\ref{fig:tools_visu}c), the same type of color-plots can be generated by calling
\functiondescription{f}{visu.\allowbreak vectorfield\_\allowbreak color\_\allowbreak by\_\allowbreak fieldindex}{}{}
The above functions are actually plotting scalar fields, the function names ``vectorfield\dots'' refer to the fact that vectorial data is taken as input.
If the data is available as scalar field (i.e. in tuples \((x,y,z,S)\) with \(S\) being a scalar value), one can use the following wrapper to \function{visu.vectorfield\_\allowbreak color}
\functiondescription{f}{visu.scalarfield}{}{}
%

%
\begin{figure}[tp]
  \centering
  \includegraphics{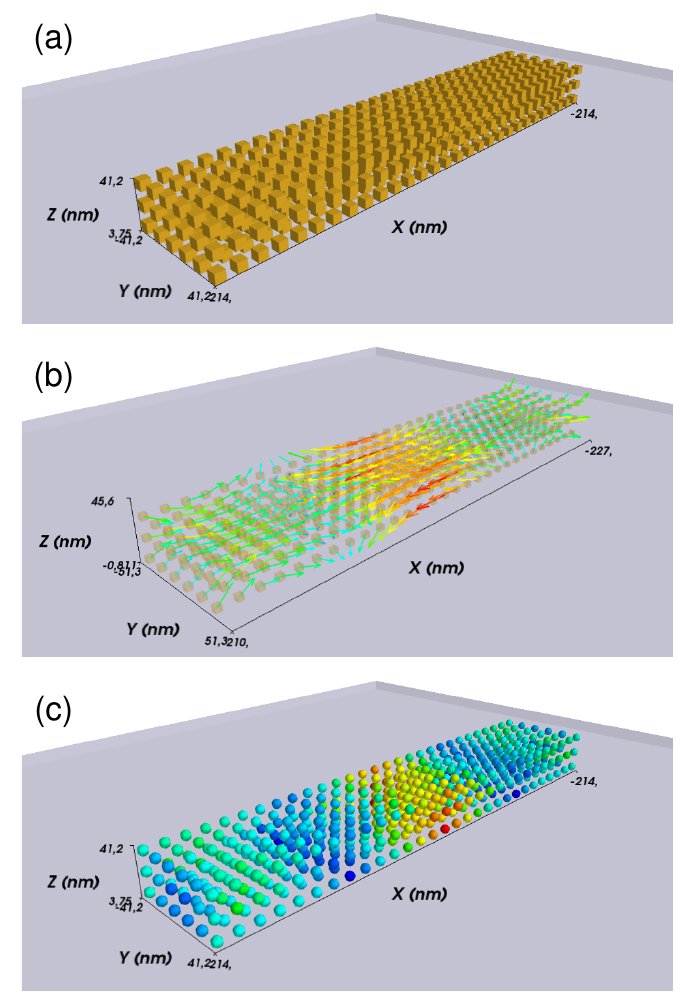}
  \caption{
  Examples illustrating the \pygdm\ 3D visualization tools on the same data as shown in figure~\ref{fig:tools_visu}a-c. (a) structure geometry, (b) electric field (real part) and (c) intensity of the electric field \(|\mathbf{E}|^2\) inside the gold nanorod.
  }\label{fig:visu3d_tools}
\end{figure}
%
%
\begin{figure*}[tp]
  \centering
  \includegraphics{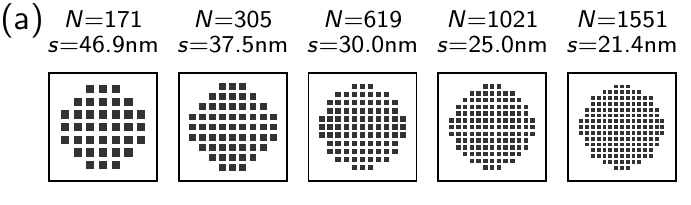}
  \hspace{.75cm}
  \includegraphics{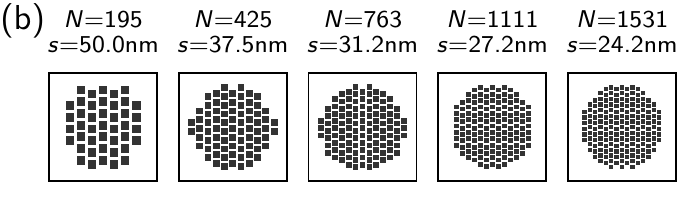}

  \includegraphics{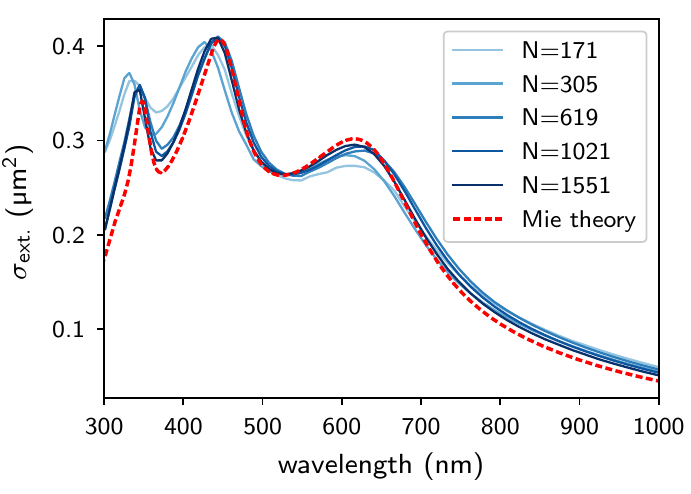}
  \hspace{.75cm}
  \includegraphics{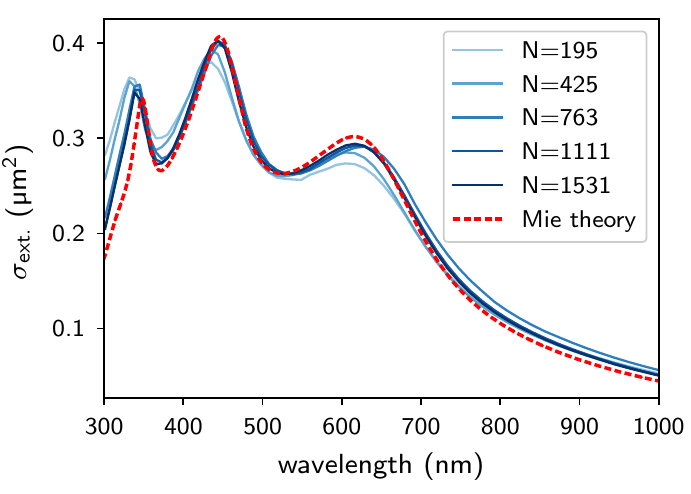}
  \caption{
  Comparison of the extinction cross-section of a dielectric sphere (\(n_{\text{sphere}}=2.0\)) of diameter \(D=300\,\)nm, placed in vacuum and illuminated by a linearly polarized plane wave. 
  Calculated either using \pygdm\ with different numbers of meshpoints (blue lines) or Mie theory (dashed red line). (a) cubic mesh, (b) hexagonal compact mesh.
  At the top the number of meshpoints \(N\), the nominal stepsize \(s\) and an illustration of the discretization are given, the latter showing \(XY\)-slices through the sphere's center. 
  }\label{fig:examples_mie_dielectric}
\end{figure*}
%

\subsubsection{Farfield backfocal plane image}

Plot the ``backfocal plane'' image scattered to the farfield from the results obtained by \function{linear.\allowbreak farfield} (see section~\ref{sec:linear_farfield})
\functiondescription{f}{visu.farfield\_\allowbreak pattern\_2D}
  { {\textit{theta}: list containing the polar angles \(\vartheta\)}
    {\textit{phi}: list containing the azimutal angles \(\varphi\)}
    {\textit{I}: list containing field intensity at  \((\vartheta, \varphi)\)}}
  {}
An example illustrating the output of the farfield plotting function is shown in figure~\ref{fig:gold_splitring_farfield}.

\subsubsection{Animate fields}

The time-dependence of time-harmonic fields is expressed by harmonic oscillations at the fixed frequency \(\omega\).
After equation~\eqref{eq:harmonicComplexWave} we can directly calculate the time-dependent field \(\mathbf{\tilde{E}} (\mathbf{r}, \omega, t)\) at time \(t\) from the complex fields \(\mathbf{E} (\mathbf{r}, \omega)\) obtained by the GDM.
\pygdm\ provides a function for simple animations of the electromagnetic fields, which allows to visualize the time-dependent optical response of nanostructures.
\functiondescription{f}{visu.animate\_\allowbreak vectorfield}
  { {\textit{NF}: list containing the complex field} }
  {}
Quiver-plots of the field vectors, the real/imaginary part of individual field components or the field intensity (as color-plots for the latter two) may be animated.

\subsection{3D visualization tools}

Similar tools as for two-dimensional data visualization are available in the \function{visu3d} module for generating 3D~figures.
The convention for the function names is the same as in the 2D visualization module in order to make switching between 2D and 3D representations as easy as possible.
Available plotting functions are \function{structure}, \function{vectorfield}, \function{vectorfield\_\allowbreak by\_\allowbreak fieldindex}, \function{vectorfield\_\allowbreak  color}, \function{vectorfield\_\allowbreak color\_\allowbreak by\_\allowbreak fieldindex} and \function{scalarfield}.
For a short explanation, see the equivalent 2D-plotting functions, described above.
Examples demonstrating the visual output of the 3D-plotting functions are shown in figure~\ref{fig:visu3d_tools} (on the same data as in figure~\ref{fig:tools_visu}a-c).

Finally, also 3D-animations of the time-harmonic fields can be generated.
This can be done using \function{visu3d.\allowbreak animate\_\allowbreak  vectorfield}.

\section{Tools}

Apart from visualization, \pygdm\ includes also several tools to render post-processing as simple as possible.

\subsection{2D-projections of nano-structures}

In order to calculate a two-dimensional projection of a nano-structure, use
\functiondescription{f}{tools.get\_geometry\_\allowbreak 2d\_projection}{}{}

\subsection{Geometric cross-section}

The geometric cross-section of a nano-structure is the area occupied by its projection onto a specific plane (i.e. its ``footprint''). 
It is often used as a reference value, for example for the scattering efficiency. 
It can be calculated using (in units of nm\(^2\))
\functiondescription{f}{tools.get\_geometric\_\allowbreak cross\_section}{}{}
By default, the projection on the \(XY\) plane is used, this can be changed via the parameter ``projection''.

\subsection{Surface of a nano-structure}

For surface-effects like surface second-harmonic generation (surface SHG), the meshpoints on the surface of a nanostructure are of particular interest.
They can be obtained using
\functiondescription{f}{tools.get\_surface\_\allowbreak meshpoints}{}{}
The function also returns the surface-normal unit vectors for each surface-meshpoint.

%
\begin{figure}[tp]
  \centering
  \includegraphics{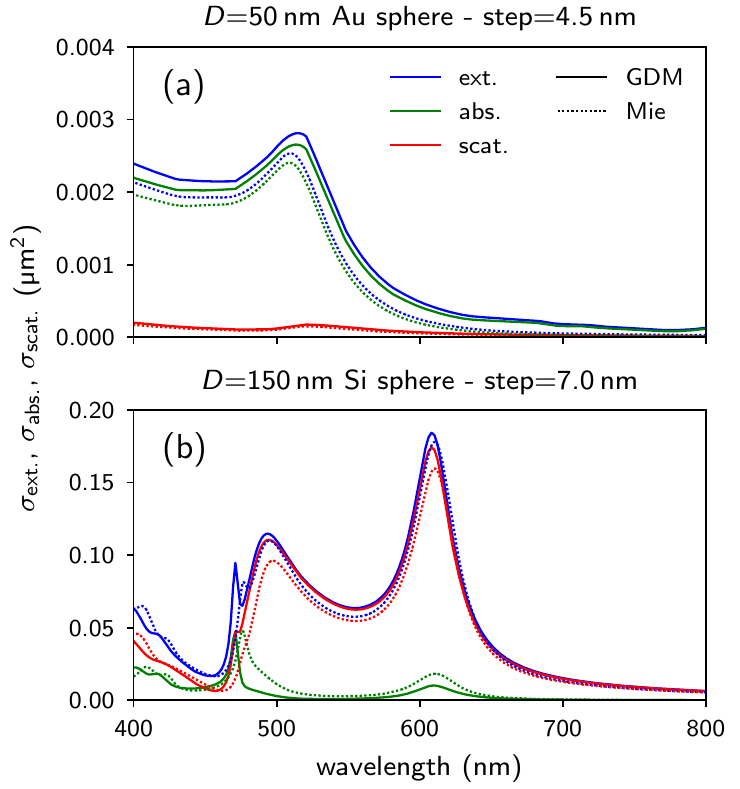}
  \caption{
  Comparison of the extinction, absorption and scattering cross-sections of (a) a gold sphere of diameter \(D=50\,\)nm and (b) a silicon sphere of diameter \(D=150\,\)nm. 
  Both spheres are placed in vacuum and illuminated by a linearly polarized plane wave. 
  Calculated either using \pygdm\ (solid lines) or by Mie theory (dotted lines).
  In both cases, a hexagonal compact mesh is used.
  }\label{fig:examples_mie_gold_silicon}
\end{figure}
%

\subsection{Calculating spectra}\label{sec:tools_spectra}

Calculating spectra of different physical quantities is a very common task in nano-optics.
\pygdm\ therefore provides tools to render this task very simple.
Each field-configurations in a \object{simulation}-object, which is available for several wavelengths, can be obtained via
\functiondescription{f}{tools.\allowbreak get\_\allowbreak possible\_\allowbreak field\_\allowbreak params\_\allowbreak spectra}{}{}
These configurations can then be used together with post-processing routines (such as \function{linear.extinct} for the extinction cross-section) to calculate a spectrum for some physical quantity.
This can be done using
\functiondescription{f}{tools.\allowbreak calculate\_\allowbreak spectrum}{}{}

\subsection{Calculating raster-scans}\label{sec:tools_rasterscan}

Because \pygdm\ uses the concept of a generalized propagator, it is particularly suited for application in monochromatic problems with varying illumination conditions, such as raster-scan simulations (varying beam position).
If a simulation with a large number of focused beam-positions has been performed, 
the available incident field configurations (e.g. wavelengths or polarizations) corresponding to full raster-scan maps can be obtained using
\functiondescription{f}{tools.\allowbreak get\_\allowbreak possible\_\allowbreak field\_params\_\allowbreak rasterscan}{}{}
Like in the case of a spectrum, a scalar mapping can be computed from a raster-scan simulation, where each raster-scan position will be attributed a value, according to an evaluation function (like \function{linear.extinct}, \function{linear.heat}, \dots).
Such maps can be obtained using
\functiondescription{f}{tools.calculate\_\allowbreak rasterscan}{}{}
%

%
\begin{figure}[tp]
  \centering
  \includegraphics{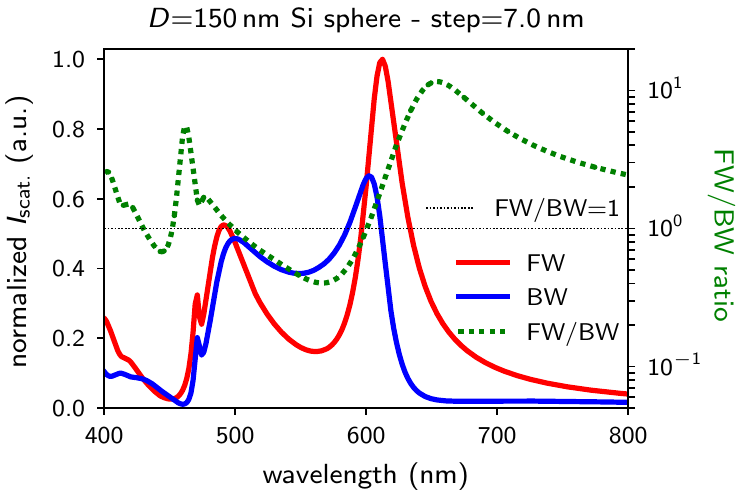}
  \caption{
  Forward (FW, red) and backward (BW, blue) scattering spectra and FW/BW ratio (green dotted) for a silicon sphere of diameter \(D=150\,\)nm in vacuum. 
  A hexagonal compact mesh is used.
  }\label{fig:examples_silicon_sphere_fw_bw}
\end{figure}
%
%
\begin{figure}[tp]
  \centering
  \includegraphics{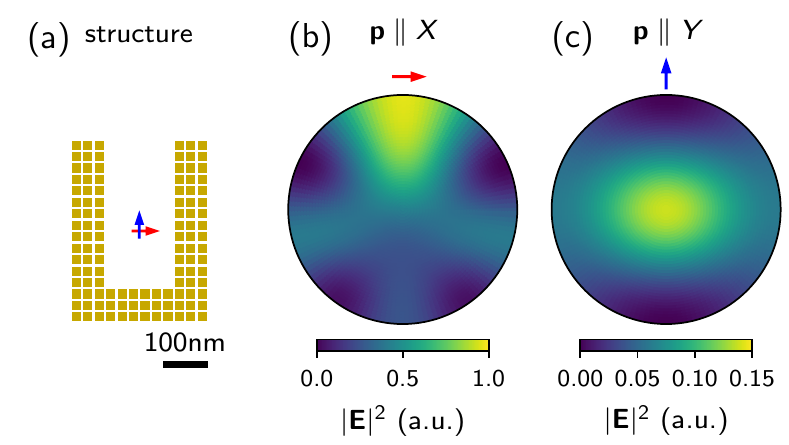}
  \caption{
  (a) sketch of the simulation geometry: A dipolar emitter, radiating at \(\lambda = 1\,\)\textmu m is placed in the center of a gold split-ring resonator (in vacuum).
  (b-c) qualitative far-field patterns (backfocal plane images) of the scattering of the quantum emitter coupled to the plasmonic structure for dipole orientations along \(X\) and \(Y\) in (b), respectively (c).
  }\label{fig:gold_splitring_farfield}
\end{figure}
%

%
\begin{figure}[tp]
  \centering
  \includegraphics{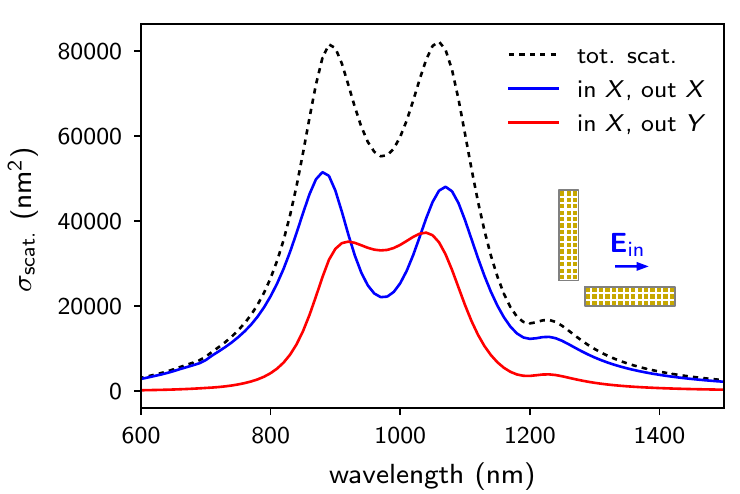}
  \caption{
  Polarization filtered scattering spectra from symmetric L-shaped plasmonic antenna (arm length \(L=210\,\)nm, width and height \(W=H=45\,\)nm).
  A sketch of the geometry is shown as inset.
  The total scattering (dashed black line) as well as the \(X\) and \(Y\) polarization filtered scattering contributions (blue and red, respectively) are shown.
  }\label{fig:examples_polarconversion}
\end{figure}
%

\section{Examples}
In the following section, we show several examples of \pygdm\ simulations. 
In the examples we try to reproduce analytical Mie theory, results from selected publications or we simply intend to demonstrate \pygdm\ features.

\subsection{Comparison to Mie theory}

Curved surfaces are generally demanding if it comes to discretization. 
A popular benchmark problem for electro-dynamical numerical methods is therefore the sphere, for which an analytical solution is given by Mie theory.
In the first examples we thus compare \pygdm\ simulations to Mie theory.

\subsubsection{Dielectric nano-sphere}

Using \function{linear.extinct}, we calculate the extinction cross-section \(\sigma_{\text{scat}}\) of a dielectric sphere of diameter \(D=300\,\)nm in vacuum with fixed, purely real refractive index \(n=2\).
Results are shown in figure~\ref{fig:examples_mie_dielectric} for different stepsizes and for (a) a cubic mesh as well as (b) a hexagonal compact lattice. 

In comparison with Mie theory, we find that the GDM offers a very good approximation already using rather coarse meshing.
Furthermore we note, that the case of a spherical particle seems to be better described using a hexagonal mesh: The agreement with the analytical solution is slightly better for comparable numbers of meshpoints.

\subsubsection{Dispersive nano-spheres (Au, Si)}

In figure~\ref{fig:examples_mie_gold_silicon} we compare spherical particles of dispersive materials. 
Fig.~\ref{fig:examples_mie_gold_silicon}a shows spectra corresponding to a \(D=50\,\)nm gold nano-sphere in vacuum, figure~\ref{fig:examples_mie_gold_silicon}b gives spectra for a \(D=150\,\)nm silicon sphere.
Simulated spectra are calculated using \function{linear.extinct} and compared to Mie theory. 
The resonance positions from Mie theory are reproduced with excellent agreement.

%
\begin{figure}[t]
  \centering
  \includegraphics{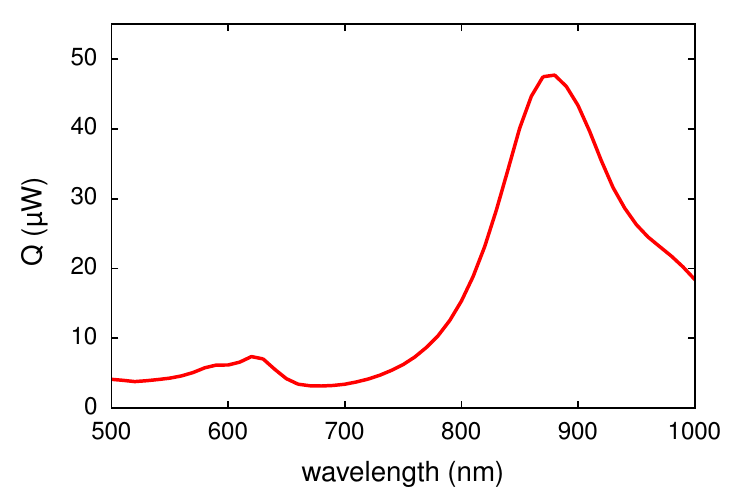}
  \caption{
  Spectrally resolved heat generation within a gold prism of side length \(L=115\,\)nm and height \(H=12\)nm. 
  Incident polarization along one edge of prism.
  }\label{fig:examples_heat_spectrum}
\end{figure}
%

%
\begin{figure*}[t]
  \centering
  \includegraphics{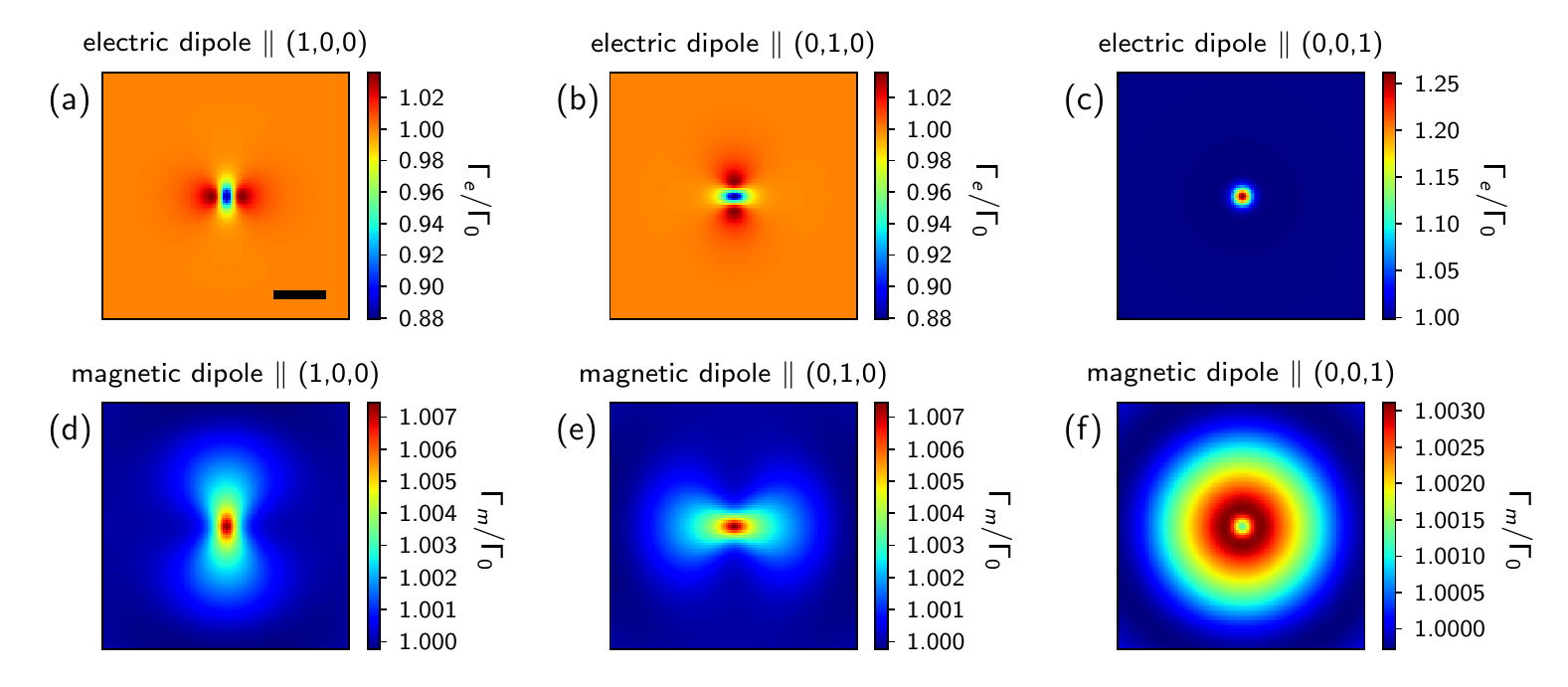}
  \caption{
  Decay rate of an electric (a-c) and a magnetic (d-f) dipole transition close to a small dielectric nano-cube (\(n=2\), side length~\(21\,\)nm, in vacuum) relative to their respective vacuum decay rate \(\Gamma_0\).
  The dipoles emit at \(\lambda_0 = 500\,\)nm, are scanned in a \(500 \times 500\,\)nm\(^2\) large plane \(15\,\)nm above the particle.
  Dipole orientations along \(0X\) (a,d), \(0Y\) (b,e) and \(0Z\) (c,f). 
  Scale bar is \(100\)\,nm.
  }\label{fig:examples_decay_rate}
\end{figure*}
%

\subsection{Other examples}

\subsubsection{Forward / backward scattering spectra}

The far-field propagation routine \function{linear.farfield} can be used to calculate directionality resolved scattering spectra.
This can be done by integrating the intensity in the far-field over limited solid angles. 
In figure~\ref{fig:examples_silicon_sphere_fw_bw}, the example of a Si sphere with diameter \(D=150\,\)nm is used again.
This time, we calculate the scattering via the \function{linear.farfield} routine (instead of using \function{linear.extinct}).
The forward (FW) and backward (BW) scattering spectra, as well as the FW/BW ratio are in excellent agreement with the results of reference \onlinecite{fu_directional_2013}.

\subsubsection{Far-field radiation pattern}

The function \function{linear.farfield} can also be used to obtain the far-field intensity distribution, comparable to experimental backfocal plane images.
In an attempt to reproduce results published in reference \onlinecite{hancu_multipolar_2014}, we put a dipolar emitter (\(\lambda_0 = 1\,\)\textmu m) in the center of a gold split-ring resonator and calculate the scattering to the far-field of the coupled system (for simplicity we consider vacuum as environment). 
The geometry of the considered arrangement is depicted in Fig.~\ref{fig:gold_splitring_farfield}a.
The dipole is oriented either along \(X\) (red) or along \(Y\) (blue), the corresponding radiation patterns are shown in figures~\ref{fig:gold_splitring_farfield}b and~c, respectively.
We can indeed reproduce the dipole-orientation dependent directionality of the scattering from the coupled system.

%
\begin{figure*}[t]
  \centering
  \includegraphics{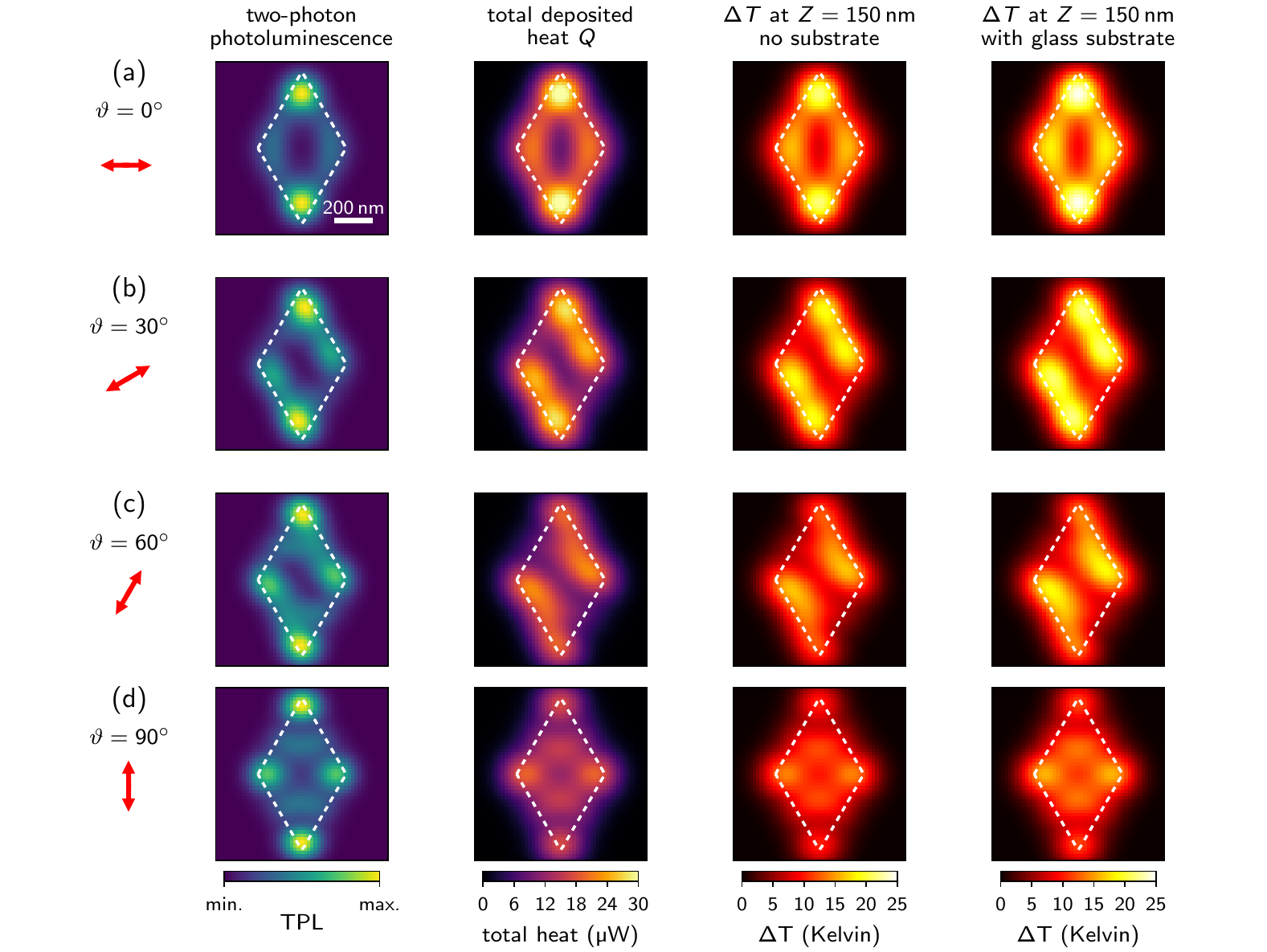}
  \caption{
  Thermoplasmonic rasterscan simulations. From left to right: TPL; total deposited heat; temperature rise \(150\,\)nm above the center of a gold rhombus (side length \(500\,\)nm, top angle \(60^{\circ}\)) as function of the focal spot position of the incident beam. 
  For the first three columns, the rhombus lies in homogeneous water. The right column shows the temperature rise for the structure in water, but lying on a glass substrate (used heat conductivities are \(\kappa_{\text{water}}=0.6\), \(\kappa_{\text{glass}}=0.8\)\,W/mK).
  The incident wavelength is \(\lambda_0 = 750\,\)nm, the linear polarization angle is (a) \(\vartheta=0^{\circ}\), (b) \(\vartheta=30^{\circ}\), (c) \(\vartheta=60^{\circ}\) and (d) \(\vartheta=90^{\circ}\).
  Scalebar in (a) is \(200\,\)nm, the position of the gold rhombus is indicated by a white dashed contour line (plotted using \function{visu.\allowbreak structure\_\allowbreak contour}).
  }\label{fig:examples_rasterscan}
\end{figure*}
%

\begin{figure*}[t]
	\centering
	\includegraphics{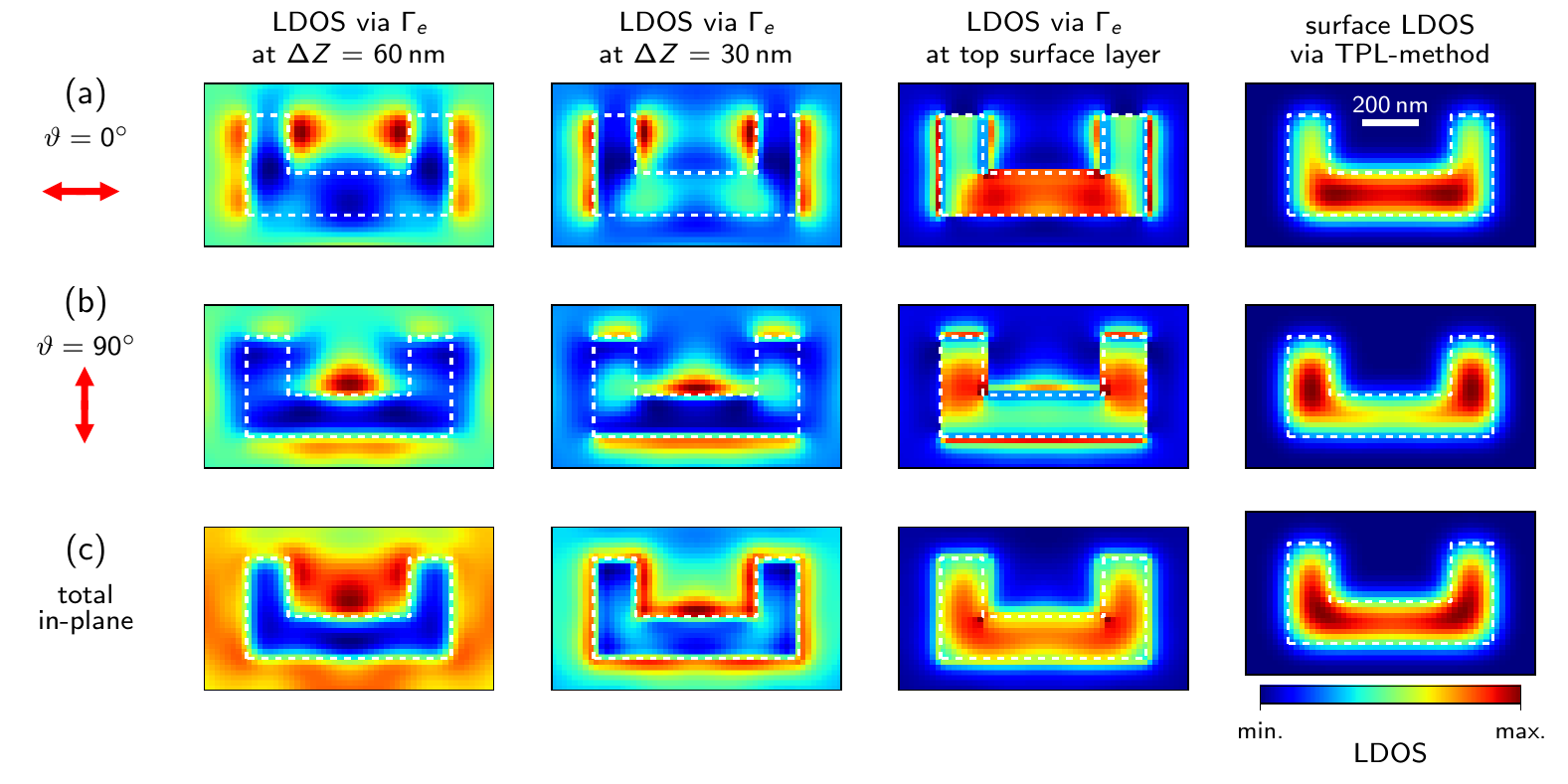}
	\caption{
		LDOS rasterscan simulations: Partial LDOS above a planar U-shaped dielectric structure (\(n=2.0\), \(H=60\,\)nm), calculated using the imaginary part of the field susceptibility (via the decay-rate of a dipolar emitter) or a raster-scanned focused beam (``TPL-method''). 
		From left to right: At \(\Delta Z = 60\,\)nm and \(\Delta Z = 30\,\)nm above the structure surface, at the height of the top-most mesh-point layer inside the structure and using the focused-beam approximation technique (see section ``TPL'').
		The wavelength is \(\lambda_0 = 600\,\)nm, the partial LDOS is shown for (a) \(\vartheta=0^{\circ}\) (\(X\)-direction), (b) \(\vartheta=90^{\circ}\)  (\(Y\)-direction) and (c) the total LDOS in the structure plane.
		The scale bar is \(200\,\)nm, the position of the structure is indicated by white dashed contour lines (plotted using \function{visu.\allowbreak structure\_\allowbreak contour}).
	}\label{fig:LDOS_example}
\end{figure*}
%

%
\begin{figure*}[t]
  \centering
  \includegraphics{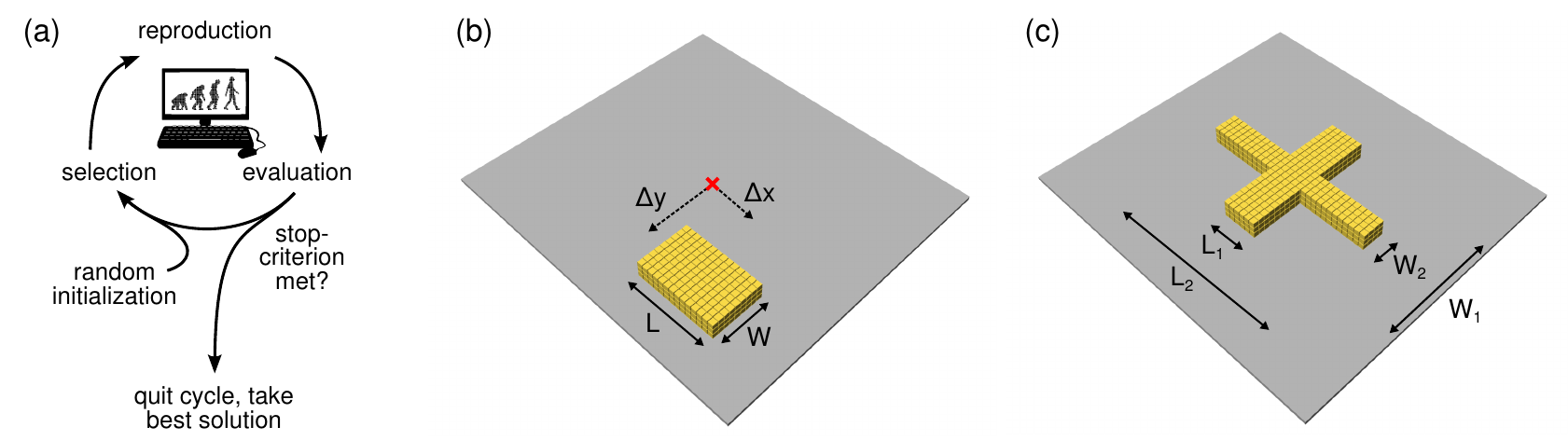}
  \caption{
  (a) Illustration of the evolutionary optimization scheme. 
  (b-c) sketches of the gold antenna geometry models used for the evolutionary optimization examples. 
  Free parameters for the rectangular geometry (b) are the length \(L\) and width \(W\) of the rectangle and an offset \((\Delta x, \Delta y)\) for the structure position with respect to the origin (indicated by a small red cross).
  Free parameters for the cross-like geometry (c) are the lengths \(L_1, L_2\) and widths \(W_1, W_2\) of the two rectangular components, forming the cross.
  }\label{fig:eo_cycle_models}
\end{figure*}
%

\subsubsection{Polarization conversion}

Also the polarization of the scattered light can be analyzed using \function{linear.farfield}.
In figure~\ref{fig:examples_polarconversion} we demonstrate polarization conversion from an L-shaped gold antenna with perpendicular arms of equal dimensions (c.f. Refs.~\onlinecite{black_optimal_2014, wiecha_polarization_2017}).
An L-shaped plasmonic antenna (in vacuum) with arm dimensions \(L=210\,\)nm, \(W=H=45\,\)nm (see inset in figure~\ref{fig:examples_polarconversion}) is illuminated by a plane wave of linear polarization along one antenna arm (here along \(X\)).
The scattered intensity is shown for two different output polarizations in blue (\(\mathbf{E}_{\text{scat}}\parallel X\)) and red (\(\mathbf{E}_{\text{scat}}\parallel Y\)), the latter corresponding to a polarization converted scattered field, which is highest if the incident wavelength is spectrally inbetween the pure modes (the pure modes correspond to polarization angles of \(\pm 45^{\circ}\), see e.g. Ref.~\onlinecite{black_optimal_2014}).

\subsubsection{Heat generation}

To demonstrate the capabilities to model nano-optical thermal effects in \pygdm, we reproduce results published in Ref.~\onlinecite{baffou_heat_2009}.
A gold prism of side length \(L=115\,\)nm and height \(H=12\,\)nm 
is illuminated by a plane wave, linearly polarized along a side of the prism.
The prism is placed on a glass substrate (\(n_{\text{subst}}=1.45\)) and is surrounded by water (\(n_{\text{env}}=1.33\)).
The total deposited heat \(Q\) from an incident power density of \(1\,\)mW\(/\)\textmu m\(^2\) is shown in figure~\ref{fig:examples_heat_spectrum} as function of the wavelength.

\subsubsection{Decay rate of dipole transition}

The modification of the decay rate of an electric and a magnetic dipolar transition close to a very small dielectric nano-particle is demonstrated in figure~\ref{fig:examples_decay_rate} (compare also with Ref.~\onlinecite{wiecha_decay_2018}).
The dipolar emitter (\(\lambda_0=500\,\)nm) is raster-scanned in the \(XY\) plane at \(\Delta z=15\,\)nm above a dielectric nano-cube (\(n=2\)) of side-length \(D = 21\,\)nm, placed in vacuum. 
At each position in the raster-scan, the relative decay rate modification with respect to the vacuum value \(\Gamma_0\) is calculated.

A noteworthy observation is the much narrower confinement of the features in the case of an electric dipole compared to the magnetic transition.
Furthermore, also the magnitude of the decay rate variation is much stronger for the electric dipole.
Both phenomena can be attributed to the more ``direct'' interaction of an electric dipole with the nano-structure, compared to the ``indirect'' magnetic response of the itself \textit{non-magnetic} nano-particle (see also section~\ref{sec:e_m_decay}).

\subsubsection{Rasterscan simulation: TPL / heat / temperature}

To demonstrate a rasterscan simulation, we calculate as function of a focused beam's focal spot position and for several incident linear polarizations: the two-photon photoluminescence (TPL) signal, the total deposited heat \(Q\) and the temperature rise at \(150\,\)nm above the center of a flat gold rhombus. 
The object is assumed to lie in water with \(n_{\text{water}}=1.33\) and a thermal conductivity of \(\kappa_{\text{water}} = 0.6\,\)W/mK.
The temperature rise is calculated either for the rhombus in a homogeneous water environment, or in water lying on a glass substrate (using \(n_{\text{glass}}=1.5\), \(\kappa_{\text{glass}}=0.8\,\)W/mK).
The rhombus dimensions are defined by a side length of \(L=500\,\)nm, a height of \(H=20\,\)nm and a top (and bottom) angle of \(60^{\circ}\).
A linearly polarized focused plane wave (\(\lambda_0=750\,\)nm) of spotsize \(w=200\,\)nm is used, setting the power density to \(1\,\)mW\(/\)\textmu m\(^2\).
The rasterscans consisting of \(50\times 50\) focal spot positions are shown in figure~\ref{fig:examples_rasterscan} for different angles of the linear polarization of the fundamental field.
We can clearly observe the correlation between TPL and the heat and temperature mappings.
We also see that the temperature rise is slightly stronger if a glass substrate is present. This is a result of heat reflection at the glass surface.

\subsubsection{Rasterscan simulation: LDOS}

In figure~\ref{fig:LDOS_example} we show rasterscan simulations of the photonic LDOS above a U-shaped dielectric planar structure. 
The length is \(800\)\,nm in the \(X\)-direction, \(400\,\)nm in the \(Y\) direction, its height is \(60\,\)nm and the bar width is \(180\,\)nm.
Fig.~\ref{fig:LDOS_example}a) shows the partial LDOS for \(X\)-oriented dipole emitters, (b) the case of \(Y\) orientation and (c) the total LDOS in the structure plane.
From left to right is shown the LDOS at decreasing distance to the structure (\(60\,\)nm, \(30\,\)nm and \(0\,\)nm to the top surface).
In the very right column the surface LDOS is calculated using the ``TPL''-method (using a spotsize of \(w=100\,\)nm, see also section~\ref{sec:nonlinear_effects}). 
Comparing the LDOS at the top surface layer with the TPL-method, the general trends are reproduced. 
The differences are not very surprising, since the calculated quantities are not exactly the same. 
The TPL method gives a measure of the energy that can be coupled into the structure at the respective surface position using a focused beam. It is therefore only non-zero if the focused beam intersects with the structure. 
The LDOS corresponds to the efficiency of the radiative coupling between a dipolar emitter and the structure and is non-zero also outside the structure.

%
\begin{figure*}[t]
  \centering
  \includegraphics{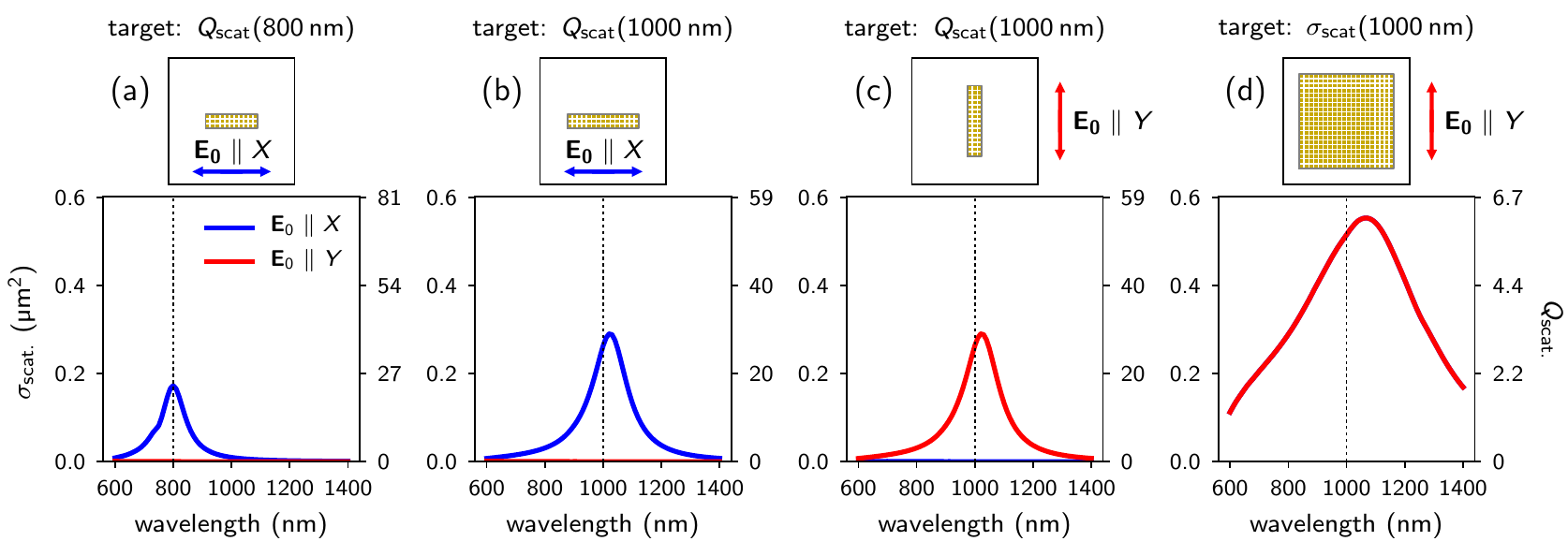}
  \caption{
  Evolutionary optimization of the dimensions of a rectangular plasmonic gold antenna for different optimization targets.
  The optimum geometry is shown in the top panels (showing \(400\times 400\,\)nm\(^2\) large areas). 
  The corresponding scattering spectra are shown in the bottom panels for plane wave illumination with \(X\) and \(Y\) linear polarization (blue, respectively red lines).
  The left ticks in each spectrum denote the scattering cross section \(\sigma_{\text{scat}}\), the ticks on the right hand side give the corresponding scattering efficiency (\(Q_{\text{scat}} = \sigma_{\text{scat}} / \sigma_{\text{geo}}\)).
  (a) maximization of \(Q_{\text{scat}}\) for \(X\) polarized illumination at \(\lambda=800\,\)nm.
  (b) maximization of \(Q_{\text{scat}}\) for \(X\) polarized illumination at \(\lambda=1000\,\)nm.
  (c) maximization of \(Q_{\text{scat}}\) for \(Y\) polarized illumination at \(\lambda=1000\,\)nm.
  (c) maximization of \(\sigma_{\text{scat}}\) for \(Y\) polarized illumination at \(\lambda=1000\,\)nm.
  }\label{fig:examples_eo_farfield}
\end{figure*}
%
%
\begin{figure}[tp]
  \centering
  \includegraphics{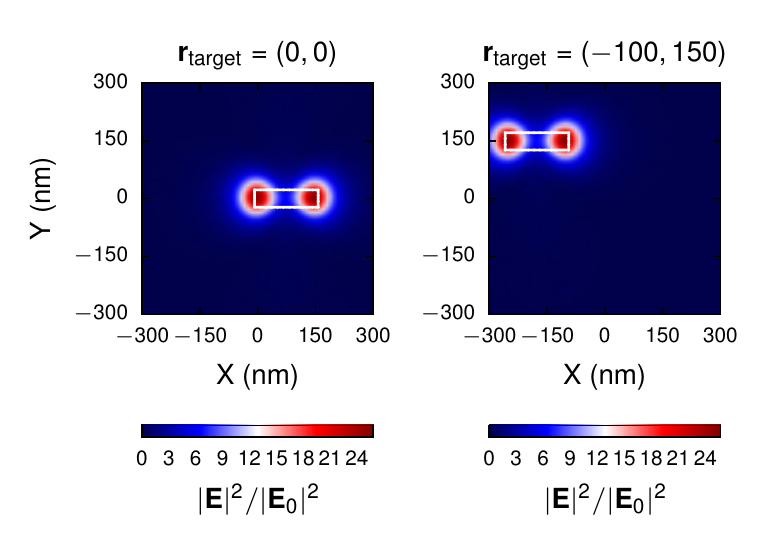}
  \caption{
  Evolutionary optimization of the dimensions and the position of a rectangular plasmonic gold antenna for maximum \(\mathbf{E}\)-field intensity enhancement at the location (a) \(\mathbf{r}_{\text{target}} = (0,0)\ [\text{nm}]\) and (b) \(\mathbf{r}_{\text{target}} = (-150,100)\ [\text{nm}]\).
  The \(z\)-coordinate of \(\mathbf{r}_{\text{target}}\) is fixed to \(30\,\)nm above the upper surface of the structure.
  Shown areas are \(600\times 600\,\)nm\(^2\).
  }\label{fig:examples_eo_nearfield}
\end{figure}
%

\section{Evolutionary optimization of nanostructure geometries}

\subsection{Evolutionary optimization}

A peculiarity of \pygdm\ is the \function{EO} module, provided together with the main \pygdm\ toolkit.
The purpose of the \function{EO} module is to find nanostructure geometries that perform a certain optical functionality in the best possible way. 
This is also known as the ``inverse problem'' \cite{macias_application_2004, odom_multiscale_2012}.
We try to achieve this goal by formulating the optical property as an optimization problem which takes the geometry of the particle as input.
Such problem will usually result in a complex, non-analytical function and hence cannot be solved by classical optimization methods like variants of the ``Newton-Raphson method''. 

In our approach we therefore optimize the problem using evolutionary optimization (EO) algorithms.
The latter mimic natural selection to find ideal solutions to complex (often non-analytical) problems. 
The initial step is to define a ``population'' of random parameter-sets for the problem.
These ``individuals'' are then evolved through a cycle of ``reproduction'' (mixing parameters between the individuals and application of random changes) and ``selection'' (problem evaluation and discarding weak solutions).
After a sufficient number of iterations, hopefully an optimum parameter-set for the problem has been found. 
The evolutionary optimization cycle is depicted in figure~\ref{fig:eo_cycle_models}a.
Unfortunately, convergence can in principle never be guaranteed in EO.
Convergence is therefore probably the most critical point in evolutionary optimization strategies. 
To ensure the credibility of the optimization results, a good stop-criterion and/or careful testing of the convergence and reproducibility of the solution for different initializations are crucial.
For details on EO, we refer to the related literature, e.g. Ref.~\onlinecite{schwefel_evolution_1993}.

\subsection{EO in \pygdm}

\pygdm\ can be used to optimize particle geometries for nano-optical problems via evolutionary optimization.
Our approach consists of three main ingredients:

\begin{enumerate}
 \item[1.] The \textbf{structure-model}: Constructs a particle geometry as function of a set of input-parameters which will be the free parameters for the optimization algorithm. It furthermore contains the simulation setup (via an instance of \object{core.simulation}). 
 It is defined by a class inherited from 
\end{enumerate}
\functiondescription{o}{EO.models.BaseModel}
  { {\textit{sim:} instance of \object{core.simulation} } }
  {}
\begin{enumerate}
 \item[2.] The \textbf{problem}: Defines the optimization target. This will usually be an optical property of the particle such as the scattering cross-section or its near-field enhancement. It is defined by a class inherited from 
\end{enumerate}
\functiondescription{o}{EO.problems.BaseProblem}
  { {\textit{sim:} instance of \object{core.simulation} } }
  {}
\begin{enumerate}
 \item[3.] The \textbf{EO algorithm}: Finally the algorithm to solve the optimization problem
\end{enumerate}

For (3.) we use the PyGMO / paGMO toolkit \cite{biscani_global_2010}.
PyGMO not only offers a large spectrum of EO algorithms.
It can furthermore distribute the population of solutions on several ``islands'' within the so-called ``Generalized island model''. 
This allows for a very easy scaling of the evolutionary optimization on multi-processor architectures \cite{izzo_generalized_2012}.

\textit{Note:} In this context, the use of the generalized propagator in \pygdm\ is a clear asset.
Optimization problems with the incident field shape and/or polarization as variable and a fixed structure geometry can be solved very efficiently. 
This includes problems like near-field shaping in adaptive optics \cite{brixner_ultrafast_2006}.

\subsection{Multi-objective optimization}

\pygdm's \function{EO}-module is also capable of treating multi-objective optimization problems, by internally addressing pyGMO's according API.
In other words, it is possible to search for nano-structure geometries optimizing \textit{multiple} target properties simultaneously.

Such evolutionary multi-objective optimization (EMO) can in principle be done in two ways: 
The first approach consists in summarizing the multiple target values in one single fitness function, hence capturing the problem into a single objective optimization.
In this case, the critical part is the construction of an appropriate fitness-value, which is usually not trivial at all.
In the second approach, one searches for the set of ``non-dominated'' or ``Pareto optimum'' solutions, which is often called the ``Pareto front''.
It consists of solutions that are all optimal in the sense that an improvement in one of the target functions necessarily leads to a decrease in at least one of the other optimization targets.
The obvious advantage is, that the individual objectives can be used \textit{as-is} without the need to fiddle them into a single fitness-function.
On the other hand, the latter approach additionally requires the selection of a single optimum solution from the set of Pareto optimum solutions.
For a detailed introduction to EMO, see e.g. Ref.~\onlinecite{deb_multi-objective_2001}.

\subsection{EO-Examples}

%
%
\begin{figure}[tp]
  \centering
  \includegraphics{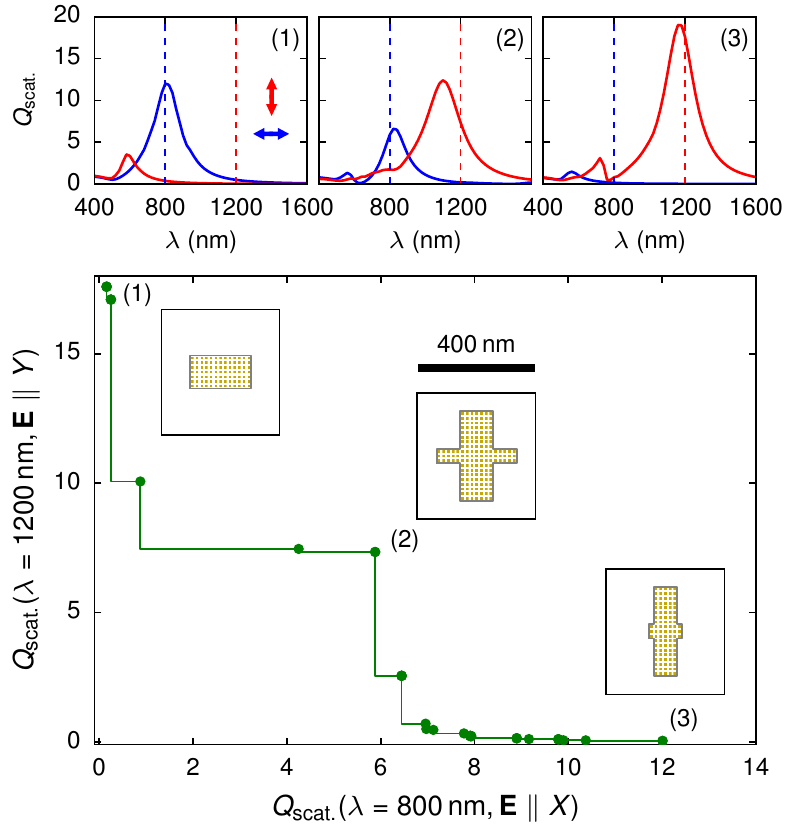}
  \caption{
  Multi-objective optimization of double-resonant plasmonic antennas made from gold.
  Large plot: Pareto front found from a concurrent maximization of \(Q_{\text{scat}}\) at \(\lambda_1=800\,\)nm and \(\lambda_2=1200\,\)nm, for polarizations along~\(X\) and~\(Y\), respectively (\(Q_{\text{scat}} = \sigma_{\text{scat}} / \sigma_{\text{geo}}\)). 
  Top: Spectra for~\(X\) (blue) and \(Y\)-polarized (red) illumination of three selected structures on the Pareto front, shown as insets (labeled by numbers 1-3).
  }\label{fig:examples_emo}
\end{figure}
%

We demonstrate the evolutionary optimization toolkit ``\function{EO}'' on some simple but illustrative problems.

\subsubsection{Maximize scattering cross-section or scattering efficiency}

For a first demonstration, we want to optimize the shape of a rectangular gold-antenna in order to obtain maximum scattering at a certain wavelength and for a fixed angle of the linear incident polarization.
The free parameters are the length \(L\) and width \(W\) of the plasmonic rectangle (see Fig.~\ref{fig:eo_cycle_models}b).
The position of the rectangle (\(\Delta x = \Delta y = 0\)), the stepsize (\(s=15\,\)nm) of the cubic mesh and the height of the antenna (\(H=45\,\)nm) are fixed.

We run the EO of the rectangular shape with different optimization targets, the respective final best solutions are shown in figure~\ref{fig:examples_eo_farfield}: 
In (a) the scattering efficiency \(Q_{\text{scat}}\) (i.e. the scattering cross-section \(\sigma_{\text{scat}}\) divided by the geometrical cross-section \(\sigma_{\text{geo}}\)) is maximized for an incident plane wave with wavelength \(\lambda_0=800\,\)nm and linear polarization of \(\mathbf{E}_0\) along \(X\).
In (b) maximum \(Q_{\text{scat}}\) is searched for \(\lambda_0=1000\,\)nm and \(\mathbf{E}_0 \parallel X\).
In (c) \(Q_{\text{scat}}\) is again maximized for \(\lambda_0=1000\,\)nm but a perpendicular polarization angle, hence \(\mathbf{E}_0 \parallel Y\).
Finally, (d) shows and optimization of the scattering cross-section \(\sigma_{\text{scat}}\) (instead of \(Q_{\text{scat}}\)), with otherwise equal configuration as in (c).

The first observation is, that the optimization is indeed capable of adjusting the size of the antenna such that the surface plasmon resonance occurs at the target wavelength.
We also observe that while the optimization of the scattering efficiency (\(Q_{\text{scat}}\) given at the right of each plot) leads to thin rectangles with low geometric cross section, the optimization of the scattering cross section (\(\sigma_{\text{scat}}\) given at the left of each plot) leads to a structure of maximum allowed dimensions. 
This leads to a cross-section \(\sigma_{\text{scat}}\) about twice as large as for the other antennas.
The scattering efficiency \(Q_{\text{scat}}\) on the other hand is significantly lower (about a factor \(5\)) compared to the optimizations shown in figure~\ref{fig:examples_eo_farfield}a-c.

\subsubsection{Maximize electric field intensity}

In a second example we want to find a structure to maximize the electric field intensity at a specific point \(\mathbf{r}_{\text{target}}\), \(30\,\)nm above the structure surface.

As structure to be optimized, we use again the rectangular gold-antenna of variable length and width with the same configuration as in the first example.
Additionally we introduce as free parameters the offsets \(\Delta x\) and \(\Delta y\), shifting the rectangle with respect to the origin of the coordinate system (see Fig.~\ref{fig:eo_cycle_models}b).
The structure is illuminated by a plane wave (\(\lambda_0=800\,\)nm), linearly polarized along \(X\).

We run the optimization for two different \(\mathbf{r}_{\text{target}}\).
The results are shown in figure~\ref{fig:examples_eo_nearfield}.
In both runs, the optimization found a plasmonic dipole antenna, resonant at the incident wavelength and shifted its position such that the hot-spot of maximum field enhancement lies at \(\mathbf{r}_{\text{target}}\).

\subsubsection{EMO: Double resonant plasmonic antenna}

In a last example, we show how multiple objectives can be optimized concurrently in a single optimization, by calculating the Pareto-front.
For the demonstration, we try to obtain structures that scatter light at two different wavelengths for perpendicular polarization angles of the incident plane wave.
We choose a simple cross-like geometry model (structure placed in vacuum), consisting of the four free parameters \(L_1, L_2\) and \(W_1, W_2\) (see Fig.~\ref{fig:eo_cycle_models}c).

The optimization goal is to simultaneously maximize \(Q_{\text{scat}}\) for (1) a wavelength \(\lambda_0 = 800\,\)nm and an incident polarization along \(X\) and (2) \(\lambda_0 = 1200\,\)nm and an incident polarization along \(Y\).
The Pareto-front obtained by the evolutionary optimization is shown in figure~\ref{fig:examples_emo}. 
Scattering spectra of selected structures are shown in the top, their geometries are illustrated in the insets, labeled (1)-(3).

The optimization indeed found structures which maximize either one of the scattering-targets ((1) and (3)), or scatter light similarly strong for both target configurations (structure (2)).
Note that the Pareto-front is not very smooth. 
In addition, the model seems not to be sufficiently sophisticated in order to obtain structures with equal \(Q_{\text{scat}}\) for both target conditions.
A more general structure model could probably provide better solutions for the problem.

\paragraph*{Note:} 
Further nano-photonic EO problems tackled using the \pygdm\ toolkit, can be found in Refs.~\onlinecite{wiecha_decay_2018, wiecha_linear_2016, wiecha_evolutionary_2017, girard_designing_2018, wiecha_multi-resonant_2018}.

\section{Conclusion}\label{sec:conclusion}

In conclusion, we presented a python toolkit for electro-dynamical simulations in nano-optics, based on a volume discretization approach, the Green dyadic method.
While other techniques like FEM may offer better accuracy, the main strength of \pygdm\ is the efficient treatment of large monochromatic problems with many illumination configurations, like raster-scan simulations. 
Such calculations can be solved very efficiently in \pygdm\ thanks to the concept of the generalized propagator. 
Furthermore, its simplicity is a great advantage of \pygdm . 
The high-level python API as well as many tools for rapid data analysis and visualization render standard nano-optical simulations very easy.
Finally, the evolutionary optimization submodule is a unique feature, which allows to optimize nanostructure geometries for specific target optical properties.
Scripts to reproduce all above shown examples, can be found online together with further, extensive documentation.

\section*{Acknowledgments}

I gratefully thank Christian Girard and Arnaud Arbouet for their advise, help with the theory, careful proof-reading and for the fortran routines. 
I also want to thank Gérard Colas des Francs for helpful discussions and his contributions to the fortran code to which also Renaud Marty contributed.
I finally thank Vincent Paillard and Aurélien Cuche for many inspiring discussions and proof-reading of the manuscript.
This work was supported by Programme Investissements d'Avenir under the program ANR-11-IDEX-0002-02, reference ANR-10-LABX-0037-NEXT, and by the computing facility center CALMIP of the University of Toulouse under grant P12167.

\section*{Conflicts of interests}

The author declares no competing financial interest.

\section{Appendix -- Accuracy, possible system size and limitations}\label{sec:appendix_accuracy}

\paragraph{Limitations}
The limit for the number \(N\) of discretization meshpoints depends mainly on the amount of RAM available in the machine. The memory requirement rises with \(3N^2\) (figure~6b).
The computation time rises even proportional to \(N^3\) (see also figure~6a), so at some point, the speed will be limiting as well.
This effectively limits the number of meshpoints to \(\approx 10000-15000\).

\paragraph{Accuracy, large systems}
To yield a reasonable accuracy, the discretization stepsize has to be sufficiently small (in the order of \(\approx 10\,\)nm for plasmonics and dielectrics of refractive index \(n \lesssim 3 \)). 
For dielectrics with higher refractive index, the discretization should be further refined to yield accurate results. 
However, if the user  is aware of the fact that the agreement will be only qualitative, approximative simulations are possible with larger discretization.
In consequence, the memory requirement limits the applicability of pyGDM to the amount of material that can be simulated with good accuracy to (very rough estimation) \(\approx 10000 \times 10^3\,\)nm\(^3\)).

\paragraph{Small systems}
When the size of the system is reduced, the discretization can be finer within the limit of the feasible \(10\)-\(15\)k meshpoints. Hence, the accuracy becomes better. One must only be aware, that pyGDM is a purely classical Maxwell solver. 
Therefore, the size of the system should not be reduced down to scales where quantum effects would occur (usually \(\lesssim 1-2\)\,nm).

\section{Appendix -- Technical details}\label{sec:technical_detail}

\subsection{Reference system in \pygdm\ simulations}

In \pygdm\ an asymptotic Green's dyad is used which describes not only a substrate (``layer 1'', refractive index \(n_1\)), but also an additional cladding layer at a variable height above the substrate (``layer 3'', \(n_3\)). 
The nanoparticle is placed in the sandwich layer (``layer 2'', \(n_2\)), i.e. in-between layers~1 and~3.
The distance between the substrate and the cladding layer can be specified by a \textit{spacing} parameter.
By default, \(n_3 = n_2\), so if the index of the cladding layer ``3'' is not specified in the constructor of \object{structures.struct}, a reference system composed of a homogeneous environment above a dielectric substrate is assumed.
To run a simulation without a substrate it is sufficient to simply set \(n_1 = n_2\).
The geometry of the \pygdm\ reference system is illustrated in figure~\ref{fig:reference_system}.

The non-retarded Dyad used in \pygdm\ can be derived using the image charges method and gives good approximations for dielectric interfaces of low refractive index.
A fully retarded Dyad for the 3-layer environment can also be calculated, which becomes necessary for instance at metallic interfaces \cite{colas_des_francs_enhanced_2005, marty_near-field_2012}.
This might be implemented in future versions of \pygdm.

\subsection{Structure geometry}

The structure geometry in \pygdm\ is defined as a list of \((x,y,z)\) tuples, defining the positions of the meshpoints on either a cubic, or a hexagonal compact, regular grid.

\begin{figure*}[tp]
  \centering
  \includegraphics[width=\linewidth]{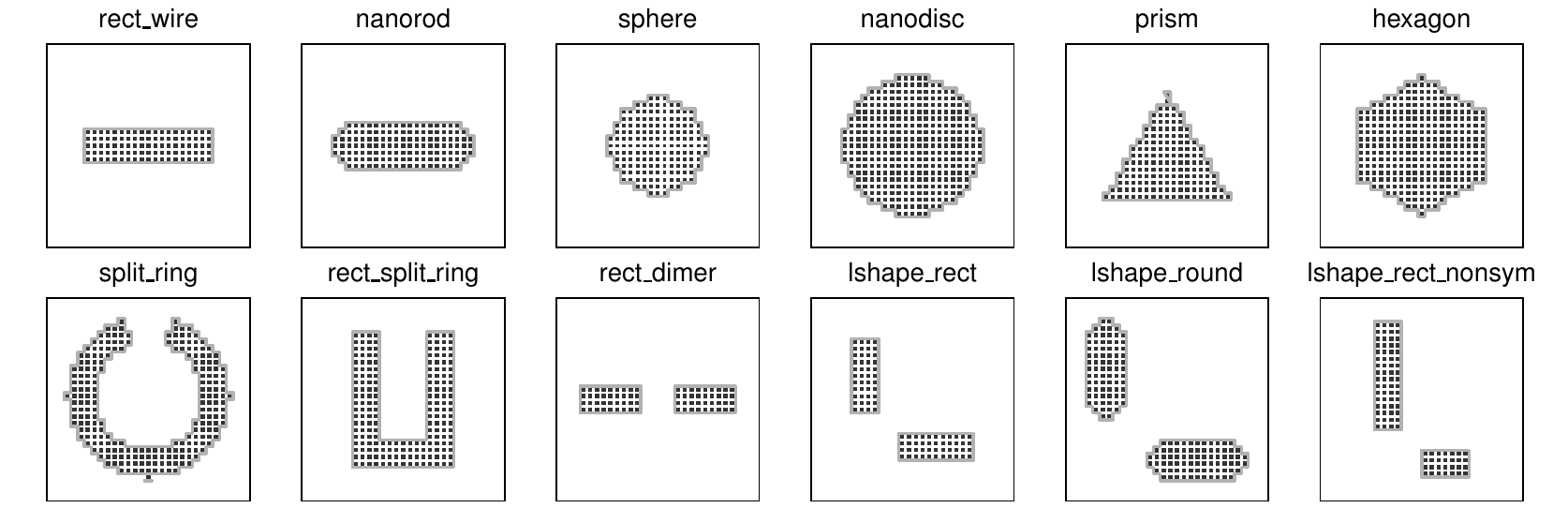}
  \caption{
  Top view of some geometries available in the \function{structures} submodule. 
  The corresponding generator function names are given on top of the example plots.
  }\label{fig:available_structures}
\end{figure*}

\pygdm\ comes with some generators for common structures in nano-photonics. These geometries are available in the \function{structures} submodule.
An overview of the available structures is shown in figure~\ref{fig:available_structures}.
Additional structures can easily be implemented at the example of the available generator functions. 
We suggest using the mesher-routines \function{structures.\_\allowbreak meshCubic} and \function{structures.\_\allowbreak meshHexagonalCompact}.

Additionally, planar structures can be generated from the brightness contrast of an image file, using
\functiondescription{f}{structures.image\_to\_struct}{}{}
This may be used to create structures from a lithography-mask layout or also from scanning electron- or atomic force-microscopy images, to simulate ``real'' geometries from an experimental sample.

Finally, structure-geometries can be manipulated (rotated, ``center of gravity'' shifted to the origin) using
\functiondescription{f}{structures.rotate\_XY}{}{}
\functiondescription{f}{structures.center\_struct}{}{}

%

\subsection{Material dispersion}

\pygdm\ provides some basic dispersion models in its \function{materials} submodule: 

\object{materials.dummy} generates a material object which returns a constant dielectric function.
``\object{silicon}'', ``\object{gold}'' and ``\object{alu}'' provide the commonly used dispersion data for the respective materials.

However, usually one would use tabulated data for the dispersion. 
This can be done in \pygdm\ via 

\functiondescription{o}{materials.fromFile}{}{}

All dispersion containers ``\object{materials.class}'' provide an \function{epsilon(wavelength)} attribute, which is a function that returns the (complex) permittivity at \textit{wavelength} (in nm).

By default, the tabulated data is interpolated linearly using \textit{numpy}'s ``\textit{interp}''.
Optionally, higher order spline interpolation is supported (based on \textit{scipy.interpolate.interp1d}).
Note that the latter may cause problems with python's ``pickling'' technique, particularly in combination with the \function{EO} module.

\subsection{Minimum working example script}

\begin{lstlisting}[language=Python, caption={Minimum example script. The plots generated by the script are shown in Fig.~\ref{fig:example_script_output}.}, label={lst:simple_example_script}]
from pyGDM2 import structures
from pyGDM2 import materials
from pyGDM2 import fields
from pyGDM2 import core
from pyGDM2 import visu

## --- simulation setup ---
## structure: sphere of 120nm radius,
## constant dielectric function (n=2),
## placed in vacuum
step = 20    # nm
geometry = structures.sphere(step, R=6, mesh='cube')
material = materials.dummy(2.0)
norm = structures.get_normalization(mesh='cube')
n1 = n2 = 1.0

struct = structures.struct(step, geometry, material, 
                                          n1,n2, norm)

## incident field: plane wave, 500nm, lin. pol. || x
field_generator = fields.planewave
wavelengths = [500]   # nm
kwargs = dict(theta=[0.0], kSign=[-1])
efield = fields.efield(field_generator, 
               wavelengths=wavelengths, kwargs=kwargs)

## create simulation object
sim = core.simulation(struct, efield)


## --- run the simulation ---
core.scatter(sim)


## --- plot the near-field inside the sphere ---
## using first (of one) field-config (=index 0)
visu.vectorfield_by_fieldindex(sim, 0, projection='XY')
visu.vectorfield_by_fieldindex(sim, 0, projection='XZ')
visu.vectorfield_by_fieldindex(sim, 0, projection='YZ')

\end{lstlisting}

\subsection{Further tools available in \pygdm}

\subsubsection{Save and load simulations}\label{sec:tools_save_load}

To save and reload pyGDM simulations, the following functions are available. Saving and loading relies on python's ``pickle'' technique:
\functiondescription{f}{tools.save\_simulation}{}{}
\functiondescription{f}{tools.load\_simulation}{}{}

\subsubsection{Show information about simulations}\label{sec:tools_print_sim_info}

To print detailed information about a \pygdm\ simulation, the following function can be used
\functiondescription{f}{tools.print\_sim\_info}{}{}
alternatively, simply use ``\textbf{print} \object{sim\_object}''.

\subsubsection{Generate coordinate list for \(2\)D map}\label{sec:tools_generate_nf_map}

To calculate \(2\)D data in \pygdm\ (e.g. near-field maps, c.f. figure~\ref{fig:tools_visu}d-g), we provide a tool to easily generate the \(2\)D grid (in cartesian \(3\)D space) for such data:
\functiondescription{f}{tools.\allowbreak generate\_\allowbreak NF\_map}{}{}

\subsubsection{Get index of specific field configuration}\label{sec:tools_get_field_index}

\pygdm\ uses keyword dictionaries to store multiple configurations of the incident field (such as several wavelengths, polarizations, focused beam positions).
All possible permutations of the given keywords are stored in the \object{core.simulation} object and are attributed an index, by which they can unambiguously identified.
In order to get the index of the field parameters that closest match specific search values (like a wavelength), one can use:

\functiondescription{f}{tools.\allowbreak get\_\allowbreak closest\_\allowbreak field\_\allowbreak index}{}{}

All field-configurations available in a simulation, sorted by their \textit{field-index}, can be obtained by

\functiondescription{f}{tools.\allowbreak get\_\allowbreak field\_\allowbreak indices}{}{}

\subsubsection{Cubic stepsize from discretized structure}

If the particle discretization is generated with another program then the \pygdm\ meshing functions (available in the \function{structures} submodule), it might be helpful to determine the stepsize of the structure.
We provide a function, that computes the stepsize of a cubic mesh by calculating the closest distance between any two meshpoints (using \textit{scipy.spatial.distance.pdist}).
\functiondescription{f}{tools.get\_step\_\allowbreak from\_geometry}{}{}

\begin{figure*}[tp]
  \centering
  \includegraphics[width=\linewidth]{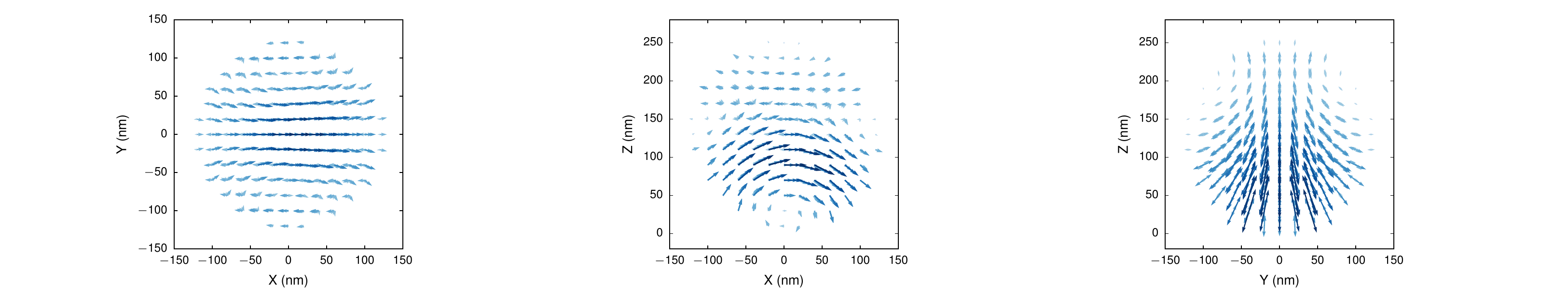}
  \caption{
  Plots generated by the demonstration script shown in listing~\ref{lst:simple_example_script}. 
  From left to right: \(XY\), \(XZ\) and \(YZ\) projections of the real part of the electric field inside a dielectric nanosphere (\(n=2\)) with radius \(R=120\,\)nm placed in vacuum. 
  Linear polarized (along \(X\)) plane wave illumination with \(\lambda=400\,\)nm, incident from positive \(Z\) (\(\mathbf{k} = -\hat{\mathbf{e}}_z k\)).
  }\label{fig:example_script_output}
\end{figure*}
%

\subsubsection{Get complex field as list of coordinate/field tuples}\label{sec:tools_get_field_as_list}

After running \function{core.scatter}, \pygdm\ stores the fields inside the particle in the \object{core.simulation} object as lists of the complex field components (\(E_{x,i}, E_{y,i}, E_{z,i}\)). 
The (\(x_i, y_i, z_i\)) geometry coordinates are stored separately in the \object{structures.struct} object within the simulation description object.
To generate complete field-lists of tuples \((x_i, y_i, z_i, E_{x,i}, E_{y,i}, E_{z,i})\), \pygdm\ provides the following functions:
\functiondescription{f}{tools.\allowbreak get\_\allowbreak field\_\allowbreak as\_\allowbreak list}{}{}
\functiondescription{f}{tools.\allowbreak get\_\allowbreak field\_\allowbreak as\_\allowbreak list\_\allowbreak by\_\allowbreak fieldindex}{}{}
Both return the complex field for a selected illumination configuration as list of coordinate / field tuples (\(x_i, y_i, z_i, E_{x,i}, E_{y,i}, E_{z,i}\)), either from the raw field-object or from the \object{simulation} object and a field-index, respectively.

\subsubsection{Generate \(2\)D map from coordinate list}\label{sec:tools_list_to_grid}

To map spatial data available as list of (coordinate-) tuples onto a plot-able 2D grid (e.g. for plotting a mapping with \textit{matplotlib.imshow}), \pygdm\ provides
\functiondescription{f}{tools.\allowbreak map\_to\_grid\_XY}{}{}

\subsubsection{Raster-scan field configurations}

If a simulation with a large number of focused beam-positions has been performed, 
the available incident field configurations corresponding to full raster-scan maps can be obtained using
\functiondescription{f}{tools.\allowbreak get\_\allowbreak possible\_\allowbreak field\_params\_\allowbreak rasterscan}{}{}
Analogously, the set of indices referring to the fields in the \object{simulation.E} which correspond to a particularly configured raster-scan, can be obtained via
\functiondescription{f}{tools.get\_\allowbreak rasterscan\_\allowbreak field\_\allowbreak indices}{}{}
Alternatively, the full set of fields inside the particle for a raster-scan with particular illumination-configuration can be obtained via
\functiondescription{f}{tools.get\_\allowbreak rasterscan\_\allowbreak fields}{}{}

\subsection{Dependencies}

The core functionalities of \pygdm\ depend only on \textbf{numpy}.
The compilation of the \textit{fortran} parts require a \textit{fortran} compiler such as \textit{gcc}'s \textbf{gfortran}.

\subsubsection{Dependencies: \function{visu}}

All 2D visualization tools require \textbf{matplotlib}.

\subsubsection{Dependencies: \function{visu3D}}

All 3D visualization tools require \textbf{mayavi}.

\subsubsection{Dependencies: \function{tools}}

Several tools require \textbf{scipy}.

\subsubsection{Dependencies: \function{structures}}

Several structure-tools require \textbf{scipy}. 
\function{image\_\allowbreak to\_\allowbreak struct} requires \textbf{PIL}.

\subsubsection{Dependencies: \function{core.scatter}: Solver (parameter ``method'')}

\pygdm\ includes wrappers to several \textbf{scipy} solvers but also other methods are supported. Below is given an exhaustive list of the available solvers and their dependencies.
For benchmarks, see figure~\ref{fig:Theory_inversion_in_GDM}.
\begin{itemize}
 \item ``lu'' (default) \textbf{scipy.linalg.lu\_factor} (LU decomposition)
 \item ``numpyinv'' \textbf{numpy.linalg.inv} (if \textbf{numpy} is compiled with LAPACK: LAPACK's ``dgesv'', else a slower fallback routine)
 \item ``dyson'': Own implementation, no requirements (sequence of Dyson's equations \cite{martin_generalized_1995})
 \item ``scipyinv'' \textbf{scipy.linalg.inv} (LAPACK's ``dgesv'')
 \item ``pinv2'': \textbf{scipy.linalg.pinv2} (singular value decomposition, SVD)
 \item ``superlu'': \textbf{scipy.sparse.linalg.splu} (superLU \cite{li_overview_2005})
 \item ``cg'': \textbf{scipy} conjugate gradient iterative solver (\textbf{scipy.sparse.linalg.bicgstab}), by default preconditioned with \textbf{scipy}'s incomplete LU decomposition (superLU \cite{li_overview_2005} via \textbf{scipy.sparse.linalg.spilu})
 \item ``pycg'': \textbf{pyamg}'s implementation of the \textit{bicgstab} algorithm, optionally preconditioned with \textbf{scipy}'s incomplete LU decomposition. Recommended if multi-threading problems are encountered with scipy's \textit{bicgstab} implementation, which is not threadsafe
\end{itemize}

\subsection{Compiling, installation}

We provide a script for the easy compilation and installation via python's ``distutils'' functionality. For this, simply run in the \pygdm\ root directory

\vspace{.5\baselineskip}
\indent\hspace{1cm}\texttt{python setup.py install}

\vspace{.5\baselineskip}
Alternatively, \pygdm\ can be compiled locally without installation via 

\vspace{.5\baselineskip}
\indent\hspace{1cm}\texttt{python setup.py build}

\vspace{.5\baselineskip}
Or it may be installed to a user-defined location using the ``\texttt{{-}{-}prefix=...}'' option.

\paragraph*{Note:} The ``setup.py'' script requires \textbf{numpy} as well as a \textit{fortran} compiler (tested with \textbf{gfortran}).

\subsection{Possible future capabilities}

The GDM can be used for manifold further calculations, which are to be included  in future versions of \pygdm. 
A non-exhaustive list of possible future features includes
\begin{itemize}
 \item 2D structures (assuming infinite length along one coordinate)\cite{paulus_greens_2001}
 \item Coherent nonlinear effects like (surface-) second or third harmonic generation\cite{wiecha_linear_2016, wiecha_origin_2016}
 \item Electron energy loss / gain spectroscopy (EELS, EEGS) or cathodoluminescence (CL) simulations\cite{geuquet_eels_2010, arbouet_electron_2014}
 \item more environment choices (surface propagator including retardation effects\cite{girard_generation_1995,girard_optical_1997}, multi-layer stratified environments\cite{paulus_accurate_2000, colas_des_francs_enhanced_2005} or magnetic decay rate calculation including a substrate\cite{girard_optical_1997, kwadrin_probing_2013})
 \item cuboidal\cite{ould_agha_near-field_2014} or non-regular meshes\cite{kottmann_accurate_2000}
 \item materials with anisotropic susceptibility, e.g. birefringent media
 \item periodic structures\cite{gallinet_electromagnetic_2009, chaumet_simulation_2009}
 \item quantum corrected model for plasmonic tunneling currents via junctions of inhomogeneous permittivity \cite{esteban_bridging_2012}
 \item SNOM image calculation/interpretation\cite{greffet_image_1997, porto_theory_2000}
 \item memory-efficient conjugate gradients solver including FFT-accelerated matrix-vector multiplications for large problems \cite{goodman_application_1991}
\end{itemize}

\section{Appendix -- Keyword arguments of the most important classes and functions}\label{sec:kwargs}

\section*{Most important classes}

For a detailed explanation of the physical information contained by the below classes, see section~\ref{sec:setup_simulation}.

\functiondescription{o}{core.simulation}{}{
 {\textit{struct:} instance of \object{structures.struct}}
 {\textit{efield:} instance of \object{fields.efield}}
}

\functiondescription{o}{structures.struct}{}{
 {\textit{step:} discretization stepsize (in nm)}
 {\textit{geometry:} list of meshpoint coordinates \((x,y,z)\) (in nm)}
 {\textit{material:} structure material dispersion, instance of \object{materials.CLASS}}
 {\textit{n1, n2:} ref. index of substrate (n1) and environment (n2)}
 {\textit{normalization (optional):} mesh-type dependent factor, default: ``1'' (cubic mesh)}
 {\textit{n3 (optional):} ref. index of cladding}
 {\textit{spacing (optional):} distance between substrate and cladding. default: ``5000'' (nm)}
}

\functiondescription{o}{fields.efield}{}{
 {\textit{field\_generator:} field generator function (e.g. from \function{fields} module)}
 {\textit{wavelengths:} list of wavelengths at which to do the simulation (in nm)}
 {\textit{kwargs (optional):} dict (or list of dict) with further kwargs for the field generator}
}

\section*{Most important functions}

For a detailed explanation of the calculations performed by the below functions, see sections~\ref{sec:solver}-\ref{sec:visu}.

\subsection*{\pygdm\ core}

\functiondescription{f}{core.scatter / core.scatter\_mpi}{}{
 {\textit{sim:} instance of \object{core.simulation}}
 {\textit{method (optional):} inversion method, default: ``lu''}
 {\textit{multithreaded (optional):} default: ``True''}
 }

\functiondescription{f}{core.decay\_rate}{}{
	{\textit{sim:} instance of \object{core.simulation}}
	{\textit{method (optional):} inversion method, default: ``lu''}
}

\subsection*{Post-processing}

\functiondescription{f}{linear.\allowbreak extinct}{}{
 {\textit{sim}: instance of \object{core.simulation}}
 {\textit{field\_index}: index of field-configuration}
}

\functiondescription{f}{linear.\allowbreak nearfield}{}{
 {\textit{sim}: instance of \object{core.simulation}}
 {\textit{field\_index}: index of field-configuration}
 {\textit{r\_probe }: list of \((x,y,z)\) coordinates at which to evaluate the near-field} 
}

\subsection*{Visualization}

\functiondescription{f}{visu.structure}{}{
 {\textit{sim}: instance of \object{core.simulation}}
 {\textit{projection (optional)}: default: ``XY''}
 {\textit{color (optional)}: optional, matplotlib compatible color, default: ``auto''}
 {\textit{scale (optional)}: scaling, default: ``0.5''}
}

\functiondescription{f}{visu.vectorfield}{}{
 {\textit{NF}: list containing the complex field (list of 6-tuples \((x_i, y_i, z_i, E_{x,i}, E_{y,i}, E_{z,i})\). See also section~\ref{sec:tools_get_field_as_list}: \function{tools.\allowbreak get\_\allowbreak field\_\allowbreak as\_\allowbreak list})}
 {\textit{projection (optional)}: default: ``XY''}
 {\textit{slice\_level (optional)}: using only fields at specific height. default: ``none'' \(\rightarrow\) superpose all vectors}
}

\functiondescription{f}{visu.scalarfield}{}{
 {\textit{NF}: list of 4-tuples containing the coordinates and scalar-field values (\((x_i, y_i, z_i, S_{i})\))}
}
%

\section{Appendix -- GDM in the SI unit system}\label{sec:SIunits}

\noindent
In order to facilitate the conversion between SI and cgs unit systems, in this section, we introduce the main GDM equations in SI units.

The Fourier transformed Maxwell equations are then:
\begin{subequations}\label{eq:Maxwell_SI}
	\begin{align}
	\nabla\cdot \mathbf{D}(\mathbf{r}, \omega)          &= \rho(\mathbf{r}, \omega)         \label{eq:MaxwellFourierdivD_SI}\\
	\nabla \times \mathbf{E}(\mathbf{r}, \omega)          &= \iu \omega \mathbf{B}(\mathbf{r}, \omega)          \label{eq:MaxwellFourierrotE_SI}\\
	\nabla\cdot \mathbf{B}(\mathbf{r}, \omega)          &= 0            \label{eq:MaxwellFourierdivB_SI}\\
	\nabla \times \mathbf{H}(\mathbf{r}, \omega)          &= -\iu \omega \mathbf{D}(\mathbf{r}, \omega) + \mathbf{j}(\mathbf{r}, \omega)    \label{eq:MaxwellFourierrotH_SI}
	\end{align}
\end{subequations}
From which the following wave equation can be derived:
\begin{equation}
(\Delta + k^2) \mathbf{E} = - \frac{1}{\epsilon_0 \epsilon_{\text{env}}}
\left(k^2  + \nabla \nabla \right) \mathbf{P}.
\label{eq:waveequationEfield_SI}
\end{equation}
This leads to the vectorial Lippmann-Schwinger equation in SI units (here for a vacuum reference system):
\begin{equation}
\mathbf{E}(\mathbf{r}, \omega)  = \mathbf{E}_0(\mathbf{r}, \omega) + \int \mathbf{G}_0^{\text{EE}}(\mathbf{r}, \mathbf{r'}, \omega) \cdot \boldsymbol{\chi}_e \cdot \mathbf{E}(\mathbf{r'}, \omega) \text{d} \mathbf{r'} 
\label{eq:LippmannSchwingerG0_SI}
\end{equation}
with the Green's Dyad
\begin{multline}\label{eq:vacuumGreenDyadicFunction_SI}
\mathbf{G}_0^{\text{EE}}(\mathbf{r}, \mathbf{r'}, \omega) 
= \frac{\e^{\iu k R}}{4\pi\epsilon_0\epsilon_{\text{env}}} \,
\Big( -k^2 \mathbf{T}_1(\mathbf{R}) \\
- ik \mathbf{T}_2(\mathbf{R}) 
+ \mathbf{T}_3(\mathbf{R}) \Big) \, ,
\end{multline}
where the definitions of \(\mathbf{T}_1\), \(\mathbf{T}_2\) and \(\mathbf{T}_3\) given in equations~\eqref{eq:vacuumGreenDyadicFunctionT1}-\eqref{eq:vacuumGreenDyadicFunctionT3} are still valid.

In Eq.~\eqref{eq:LippmannSchwingerG0_SI} and its volume discretization
\begin{multline}
\mathbf{E}(\mathbf{r}_i, \omega) = 
\mathbf{E}_0(\mathbf{r}_i, \omega) + \\
\chi_e \sum\limits_{j=1}^{N} 
\mathbf{G}_0^{\text{EE}}(\mathbf{r}_i, \mathbf{r}_j, \omega) \cdot 
\mathbf{E}(\mathbf{r}_j, \omega) V_{\text{cell}}
\label{eq:LippmannSchwingerVolumeDiscretization_SI}
\end{multline}
one has to use the susceptibility \(\chi_{\text{e}} = (\epsilon_r - \epsilon_{\text{env}})\). For simplicity here we assumed a scalar \(\chi_e\).

Finally, the renormalization tensors Eqs.~\eqref{eq:renormalization_cube} and~\eqref{eq:renormalization_hex} have to be divided by a factor \(4\pi\).

\subsection*{Post-processing routines}
Concerning the post-processing routines, some of the pre-factors need to be adapted. 
For instance, the factor before the sums of equations~\eqref{eq:sigma_ext_from_nearfield} and~\eqref{eq:sigma_abs_from_nearfield} writes in SI units: 

\begin{equation}
\frac{2\pi n}{\lambda_0 |\mathbf{E}_0|^2}
\end{equation}
with the refractive index \(n\).

The equations \eqref{eq:heat_deposited} and \eqref{eq:temp_rise_vicinity} for heat-generation, respectively the local temperature increase have to be multiplied by a factor \(4\pi\).
\paragraph*{Note:} In \pygdm\ the post-processing routines internally convert the results to SI compatible units. The ``\function{extinct}'' function for instance returns the cross sections in units of nm\(^2\), ``\function{heat}'' returns nano Watts and ``\function{temperature}'' returns \(^{\circ}\)K.
For the respective returned units, see the technical documentation of the routines in the online documentation of the API e.g. at~\href{https://wiechapeter.gitlab.io/pyGDM2-doc/apidoc.html}{https://\allowbreak wiechapeter.\allowbreak gitlab.io/\allowbreak pyGDM2-doc/\allowbreak apidoc.\allowbreak html}.

\begin{figure*}[t]
  \centering
  \includegraphics[rotate=270, scale=1.6]{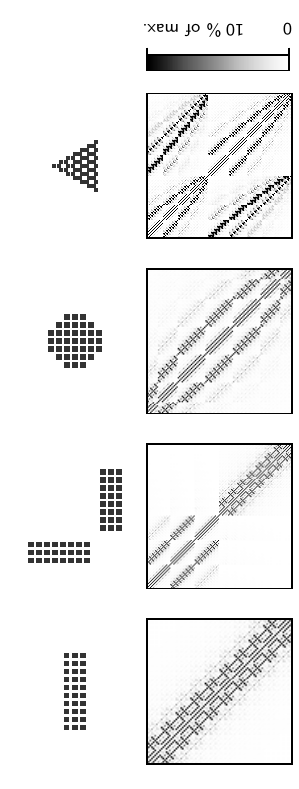}
  \caption{
  Population patterns of matrices \(\textbf{M}\) at \(\lambda=1\,\)\textmu m for a selection of structures (stepsize \(10\,\)nm, same scale for all sketches). Cubic meshes for the first three structures, hexagonal compact mesh for the structure on the right. For illustrative purposes the structures are only one layer of mesh-points high (small matrix size). White corresponds to an absolute value of \(0\), black to \(\geq 10\)\,\% of the matrix's largest element.
  }\label{fig:Theory_sparsity_examples}
\end{figure*}

\section{Appendix -- Conjugate gradients}\label{sec:conjugate_gradients}

The following applies to all \pygdm\ functions which solve the main GDM inversion problem. 
The conjugate gradients solver provides an alternative to full inversion of the coupled dipole problem which can -- under circumstances -- be preferable to complete matrix inversion.

\begin{itemize}
 \item argument \textit{method}: ``cg'' (requires \textit{scipy}) or ``pycg'' (requires \textit{pyamg})
\end{itemize}

If we have a closer look at the matrix \(\textbf{M}\) (see Eq.~\eqref{eq:definitionMforInversion}), we can make an interesting observation:
While \(\textbf{M}\) is not exactly sparse, most of the entries have significantly smaller absolute values than very few large matrix elements.
In Fig.~\ref{fig:Theory_sparsity_examples} we show plots of the population of matrix \(\mathbf{M}\) for some selected nano-structures. 
These population plots work as illustrated in the following examples:
\begin{equation*}
 \left[
 \begin{matrix}  
    1 & 0 & 0 \\ 
    0 & 1 & 0 \\ 
    0 & 0 & 1 
 \end{matrix}
 \right]
 =
 \centering\includegraphics[raise=-0.4\height]{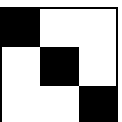}
\end{equation*}
\begin{equation*}
 \left[
 \begin{matrix}  
    2 & 1 & 0 \\ 
    1 & 2 & 1 \\ 
    0 & 1 & 2 
 \end{matrix}
 \right]
 =
 \centering\includegraphics[raise=-0.4\height]{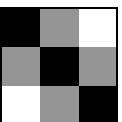}
\end{equation*}
\begin{equation*}
 \left[
 \begin{matrix}  
    1 & 2 & 3 \\ 
    4 & 5 & 6 \\ 
    7 & 8 & 9 
 \end{matrix}
 \right]
 =
 \centering\includegraphics[raise=-0.4\height]{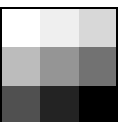}
\end{equation*}
\(\mathbf{M}\) contains also phase-information and is therefore complex, hence we use the absolute values of the matrix elements for the population patterns.
In addition, the maximum of the color-code in Fig.~\ref{fig:Theory_sparsity_examples} is clipped to 10\,\% of the maximum absolute value in the matrix to increase the contrast. 
Clearly, the matrices contain very few entries with values of more than some \% of the overall maximum and yet~\(>60\,\)\% of all elements are generally non-zero.

It turns out, that such matrices are good candidates for iterative solving using so-called ``Krylow-subspace methods''.
The most popular algorithm of this class is the conjugate gradients (CG) method and its derivations like biconjugate gradients (for non-symmetric problems) or complex CG \cite{joly_complex_1993}. 
A detailed description of the method can be found in Ref.~\onlinecite{press_numerical_2007} (chapter~2.7).
The main idea of these iterative methods is, that the inverse of the matrix is in many cases not actually required. 
For simulations that massively make use of the generalized propagator (like raster-scan simulations), the CG technique is therefore not the method of choice.
It may be on the other hand an advantageous approach, if we search a solution for \(\mathbf{E}(\omega)\) that satisfies
\begin{equation}
 \mathbf{M}(\omega) \cdot \mathbf{E}(\omega) = \mathbf{E}_0(\omega)
\end{equation}
for one single or only few incident field \(\mathbf{E}_0(\omega)\). 
During the CG-iterations, matrix-vector multiplications \(\textbf{M}\cdot\textbf{x}\) are performed following a minimization scheme in which \(\textbf{M}\cdot\textbf{x}\) converges eventually to \(\mathbf{E}_0\).
Theoretically, for a \(N\times N\) matrix CG converge to the exact solution after \(N\) iterations and each iteration itself has a computational cost \(\propto N^2\).
In reality, the convergence is often very rapid in the beginning, and a solution with sufficient precision can be obtained after very few iterations, yielding a total computational cost \(\propto N^2\) instead of a \(N^3\) scaling for exact inversion for example with LU-decomposition.
Indeed, we find a \(N^3\)-scaling for complete inversion by LU or Dyson's sequence and a \(N^2 \) dependence when using conjugate gradients (Fig.~\ref{fig:Theory_inversion_in_GDM}a). 
Particularly for larger numbers of meshpoints, this allows a reduction of the simulation time, as shown in Fig.~\ref{fig:Theory_inversion_in_GDM}a.

%
\begin{figure}[t]
  \centering
  \includegraphics{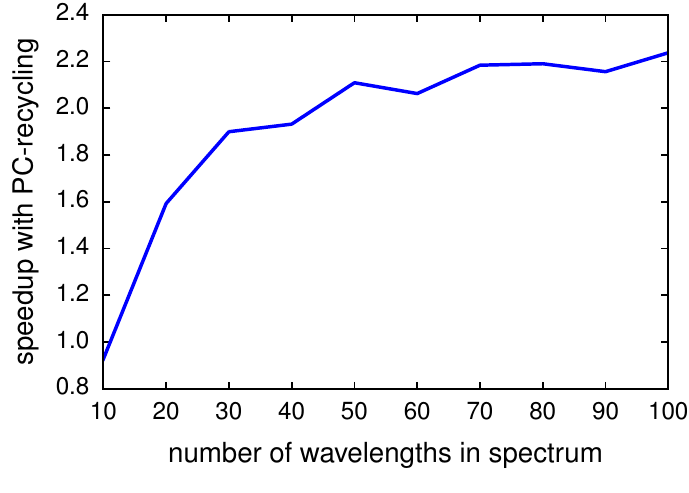}
  \caption{
  Speedup of the GDM-calculation of a spectrum (2000 meshpoints Si nanowire, step of \(10\,\)nm, \(\lambda\) from \(500\,\)nm to \(1500\,\)nm) as function of the number of wavelengths, if recycling of the preconditioner is enabled. The narrower the wavelengths in the spectrum, the higher the possible gain of PC-recycling.}\label{fig:PC_recycling_speedup}
\end{figure}
%

\subsection{Preconditioning}

\begin{itemize}
 \item argument \textit{pc\_method}: ``ilu'', ``lu'' (both require \textit{scipy}), ``amg'' (requires \textit{pyamg}) or ``none'' (no preconditioning)
\end{itemize}

The speed of the convergence of conjugate gradients is crucially dependent on the condition of the matrix \(\mathbf{M}\) and generally can be massively improved by doing a \emph{preconditioning} step before starting the actual iterative scheme. 
Let's assume, \(\mathbf{A}\) of the equation system
\begin{equation}
 \mathbf{A} \cdot \mathbf{x} = \mathbf{b}
\end{equation}
would be the identity matrix \(\mathbf{I}\). Then CG would have converged within the first iteration.
A possible approach for preconditioning is therefore to reshape the problem using a matrix \(\mathbf{P}\)
\begin{equation}\label{eq:rightHand_Preconditioning}
 \mathbf{A} \cdot \left(\mathbf{P} \cdot \mathbf{\hat{x}}\right) = \mathbf{b}.
\end{equation}
If \(\mathbf{P}\) is a close approximation to \(\mathbf{A}^{-1}\), \(\mathbf{A} \cdot \mathbf{P}\) will be close to the identity \(\mathbf{I}\) and the system would converge very quickly under conjugate gradients iterations. Eq.~\eqref{eq:rightHand_Preconditioning} is called a right-preconditioned system.
Consequently, a good preconditioner for our problem is a close approximation to the inverse of \(\mathbf{M}\). 
Several algorithms exist to search pseudo-inverse matrices for preconditioning. 
A very popular one is the \emph{incomplete} LU-decomposition (ILU) \cite{li_supernodal_2011} that scales with \(N^2\) and which is the default method in \pygdm.

\subsection{Preconditioner recycling}

\begin{itemize}
 \item argument \textit{cg\_recycle\_pc}: ``True'' (=default)
\end{itemize}

When calculating spectra using the GDM, the electric field in a particle is usually calculated for a large number of closely spaced wavelengths, at each of which the matrix \(\mathbf{M}\) is (incompletely) inverted.
Most often, the electric field distribution changes only marginally for slightly different wavelengths and so does the matrix \(\mathbf{M}\).
Unfortunately, a \emph{very similar} matrix is of little use for exact calculations, but we have seen in the preceding section that an \emph{approximation} to the exact inverse \(\mathbf{M}^{-1}\) can be a good preconditioner \(\mathbf{P}\) for CG.

When calculating dense spectra (i.e. many points on the wavelength axis), we can use this fact and significantly accelerate the calculation with conjugate gradients by recycling the preconditioner matrix until a certain lower limit for the speedup factor is reached. 
In other words, we will be using the same \(\mathbf{P}\) repeatedly for several consecutive wavelengths and only if the acceleration is below a speed-up limit, a new preconditioner is calculated and subsequently re-used for the following wavelengths.
As shown in Fig.~\ref{fig:PC_recycling_speedup}, this technique can divide the total calculation time easily by more than a factor~\(2\). 

Another possible application when preconditioner recycling may be beneficial is in series of simulations with many very similar or slowly transformed nano-structures like antennas of gradually increasing size.

\paragraph*{Note:}

The conjugate gradients solver is not very efficient for the moment and will be improved in future versions of \pygdm. 
In particular, in the specific case of the coupled dipole approximation it is possible to do very efficient vector/matrix multiplications by applying a fast Fourier transformation (FFT) scheme \cite{goodman_application_1991}.
This is not yet implemented in \pygdm.
The currently used third-party sparse-matrix solvers are not ideally suited regarding the dense matrix problem in \pygdm.



\end{document}